\documentclass[preprint,12pt]{elsarticle}

\usepackage{graphicx}
\usepackage{caption}
\usepackage{subcaption}
\usepackage{float}
\usepackage[export]{adjustbox}

\usepackage{amssymb}
\usepackage{textcomp}
\usepackage{amsmath}

\usepackage{lineno}
\usepackage{cancel}
\usepackage{color}
\usepackage{comment}
\let\oldequation\equation
\let\oldendequation\endequation

\renewenvironment{equation}
  {\linenomathNonumbers\oldequation}
  {\oldendequation\endlinenomath}

\journal{arXiv}

\begin{document}

\begin{frontmatter}



\title{A modular apparatus for use in high-precision measurements of parity violation in polarized eV neutron transmission} 


\author{D. C. Schaper$^{1,2}$, C. Auton$^{3}$, L. Barr\'on-Palos$^{4}$, M. Borrego$^{2}$, A. Chavez$^{2}$, L. Cole$^{1}$, C. B. Crawford$^{1}$, J. Curole$^{3}$, H. Dhahri$^{1}$, K. A. Dickerson$^{3}$, J. Doskow$^{3}$, W. Fox$^{3}$, M.H. Gervais$^{1}$, B. M. Goodson$^{5}$, K. Knickerbocker$^{2}$, C. Jiang$^{6}$, P. M. King$^{7}$, H. Lu$^{3}$, M. Mocko$^{2}$, D. Olivera-Velarde$^{8}$, J. G. Otero Munoz$^{3}$, S.I. Penttil\"{a}$^{6}$, A. P\'erez-Mart\'in$^{4}$, B. Short$^{9}$, W. M. Snow$^{3}$, K. Steffen$^{3}$, J. Vanderwerp$^{3}$, G. Visser$^{3}$}

\address{$^{1}$University of Kentucky, Lexington, KY, 40506, USA}
\address{$^{2}$Los Alamos National Laboratory, Los Alamos, NM, 87545, USA}
\address{$^{3}$Indiana University, Bloomington, IN, 47405, USA}
\address{$^{4}$Instituto de F\'isica, Universidad Nacional Aut\'onoma de M\'exico, Apartado Postal 20-364, 01000, M\'exico}
\address{$^{5}$Southern Illinois University, Carbondale, IL 62901, USA}
\address{$^{6}$Oak Ridge National Laboratory, Oak Ridge, TN, 37830, USA}
\address{$^{7}$Ohio University, Athens, OH, 45701, USA}
\address{$^{8}$Berea College, Berea, KY, 40404, USA}
\address{$^{9}$University of Minnesota Twin Cities, Minneapolis, MN, 55455}

\begin{abstract}
We describe a modular apparatus for use in parity-violation measurements in epithermal neutron-nucleus resonances with high instantaneous neutron fluxes at the Manuel Lujan Jr.\ Neutron Scattering Center at Los Alamos National Laboratory. This apparatus is designed to conduct high-precision measurements of the parity-odd transmission asymmetry of longitudinally polarized neutrons through targets containing nuclei with p-wave neutron-nucleus resonances in the 0.1-10 eV energy regime and to accommodate a future search for time reversal violation in polarized neutron transmission through polarized nuclear targets. The apparatus consists of an adjustable neutron and gamma collimation system, a \(^3\)He-$^{4}$He ion chamber neutron flux monitor, two identical cryostats for target cooling, an adiabatic eV-neutron spin flipper, a near-unit efficiency \(^6\)Li-\(^{7}\)Li scintillation detector operated in current mode, a flexible CAEN data acquisition system, and a neutron spin filter based on spin-exchange optical pumping of $^{3}$He gas. We describe the features of the apparatus design devoted to the suppression of systematic errors in parity-odd asymmetry measurements. We describe the configuration of the apparatus used to conduct a precision measurement of parity violation at the 0.7 eV p-wave resonance in $^{139}$La which employs two identical $^{139}$La targets, one to polarize the beam on the p-wave resonance using the weak interaction and one to analyze the polarization.     

\end{abstract}


\end{frontmatter}


\section{Introduction} \label{s:introduction}
\label{S:1}

Compound neutron-nucleus resonance reactions have proven to be an excellent laboratory in which to study parity violation~\cite{Mitchell1999, Mitchell2001}. The complex, many-body nature of this nuclear system with its very high level density was predicted theoretically~\cite{Sushkov1980, Sushkov1982} and confirmed by experiment~\cite{Alfimenkov1983, Alfimenkov1984} to provide a natural amplification mechanism for Parity Violating (PV) neutron-nucleus interactions. Statistical spectroscopy has been used to successfully analyze the widths of the distributions of observed parity-odd asymmetries in several heavy nuclei in the isolated resonance regime corresponding to eV-keV neutron energies~\cite{Tomsovic2000}. The present theory involving the mixing of nearby s-wave and p-wave resonances that accounts for this amplification of parity-odd amplitudes also predicts a similar amplification of Time Reversal Invariance Violating (TRIV) amplitudes~\cite{Fadeev2019}. Parity- (P) and Time- (T) odd effects in forward transmission provide null tests for time reversal invariance~\cite{Bowman2014}.  Compound nuclear systems are therefore attractive candidates for T-violation searches. With the advent of high-flux MW-class short-pulsed spallation neutron sources, sensitive tests of TRIV in neutron-nucleon interactions can be realized. 

Due to the proportionality between potential TRIV amplitudes and PV amplitudes, these future TRIV tests will need more complete and precise information on the properties of the resonances which exhibit large parity violation effects. In addition to new measurements of (n, $\gamma$) angular distributions~\cite{Okudaira2018} which can be used to determine the details of the quantum numbers of these $\ell=1$ (p-wave) resonances, it is also important to repeat previous measurements of P-odd neutron transmission asymmetries~\cite{Alfimenkov1983, Shimizu, Frankle, Szymanski1996, Skoy} with higher precision. By measuring the transmission asymmetry of longitudinally polarized neutrons through an unpolarized target, one can determine the parity-odd helicity dependence of the total cross section. The transmission asymmetry is defined as
\begin{equation}
\label{eq:hjonk}
A=\frac{Y^+-Y^-}{Y^++Y^-}=\tanh(f_nnt\sigma_pP)
\end{equation}
where \(Y^\pm\) is the neutron transmission yield for the two neutron helicities, \(f_n\) is the polarization of the neutron beam, \(\sigma_p\) is the p-wave cross section for unpolarized neutrons, \(n\) is the number density of the target nuclei, \(t\) is the target thickness, and \(P\) is the parity-odd longitudinal analyzing power, defined by its relationship to the helicity-dependent cross section, \(\sigma^\pm\):
\begin{equation}
\sigma^\pm=\sigma_p(1\pm f_nP)
\end{equation}


The two most effective and well-known methods to polarize neutron beams \cite{Williams1988} in the eV energy range are spin-dependent scattering from a polarized proton target~\cite{Lushchikov1970} and spin-dependent absorption in a polarized $^{3}$He target~\cite{Coulter}. There also exists a third, lesser-known polarization method which is at present practical only for the \(^{139}\)La nucleus. The \(^{139}\)La nucleus possesses a p-wave resonance near 0.7~eV, whose interference with nearby s-wave resonances results in a 10\% parity-odd asymmetry. This is larger than expected on dimensional grounds alone by about a factor of \(10^6\). As demonstrated by C.D. Bowman et al.~\cite{CDBowman1989} and later Yuan et al.~\cite{Yuan91} this parity violating asymmetry is large enough that one can use it to polarize the beam using the (intrinsically parity-odd) weak interaction. By transmitting such a weakly-polarized beam through a second \(^{139}\)La target, one can realize a measurement of the P-odd asymmetry in which the \(^{139}\)La targets act as both a polarizer and as a polarization analyzer, resulting in the aptly-named `Double Lanthanum' measurement method. Although the beam polarization produced on the parity-odd 0.7 eV resonance in $^{139}$La is small \cite{CDBowman1989}\cite{Yuan91}, its use in this configuration eliminates certain systematic errors associated with measuring \(P\), especially the absolute knowledge of the neutron beam polarization \(f_n\). In this paper we describe how our apparatus may be modified to accommodate setups for both the `Double Lanthanum' \(P\) measurement method as well as a traditional \(P\) measurement using a $^{3}$He neutron spin filter.

In addition to its relevance for future time reversal violation experiments, a precise determination of \(P\) for \(^{139}\)La is of potential use as a standard analyzer in eV neutron scattering and transmission experiments by allowing one to measure the absolute polarization of epithermal neutron beams containing \(\ell=1\), 0.7~eV neutrons. $^{139}$La was already used for this purpose in the measurements of P-odd asymmetries in heavy nuclei conducted by the TRIPLE collaboration~\cite{Mitchell2001}. It was also used in some pioneering measurements of the depolarization of polarized eV neutrons in transmission through magnetic fields in polarized and aligned nuclear targets to  extract information on the internal magnetic domain structure~\cite{Alfimenkov1997polarized}\cite{Alfimenkov1997energy}\cite{Alfimenkov1995}. The large P-odd asymmetry on the $0.88$ eV $\ell=1$ resonance in $^{81}$Br could also be used for this purpose. The possibility of constructing a compact eV neutron polarization analyzer at these two energies based on neutron absorption in LaBr$_{3}$ scintillating crystals is under analysis.  

In this paper we describe the apparatus and experimental setup we have realized to perform precision measurements of the parity violation, \(P\), present in the eV neutron energy range as configured for the Double Lanthanum experiment. The Los Alamos Neutron Science Center (LANSCE) beam facility is described in Section \ref{s:facility}. The mechanical apparatus used for the parity violation measurement, including the collimation, cryostats, mechanical rotation stage, and detector shieldhouse, are all described in Section \ref{s:apparatus}. The custom spin flipper constructed for this experiment and optimized for 0.7 eV neutrons is described in Section \ref{s:spinflipper}. Details of the electronics design for the spin flipper are described in Section \ref{s:electronics}. The fast-response current mode \(^6\)Li-$^{7}$Li glass scintillator detector designed for this experiment is discussed in Section \ref{s:detector}.
Section \ref{s:daq} describes the data acquisition system, including all measured parameters and their experimental relevance. Section~\ref{s:3He} describes the design of the polarized $^{3}$He neutron spin filter for eV neutrons to be installed for future parity violation measurements. A brief summary and description of possible future measurements which can be conducted with this instrument is given in Section \ref{s:summary}.  

\section{The FP12 neutron beam facility at LANSCE} \label{s:facility}

\begin{figure}[H]
\centering
  \includegraphics[width=\linewidth]{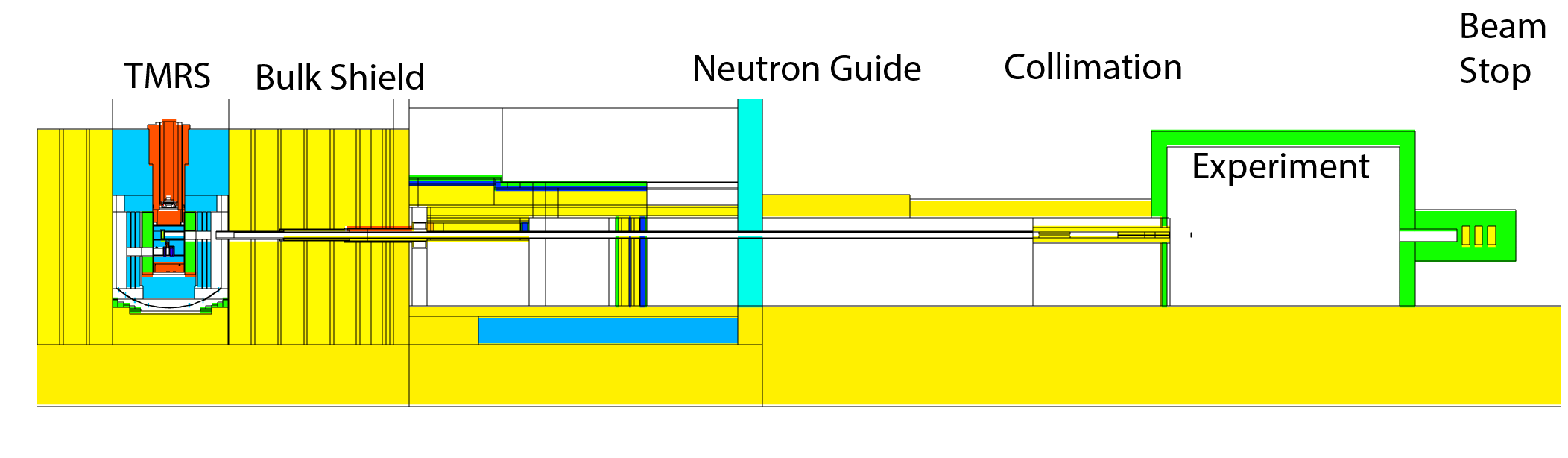}
 \caption{Elevation view of the FP12 geometry as implemented in the MCNPX model.}
  \label{figure:mcnp}
\end{figure}

The apparatus described in this paper was located in an experimental hutch on Flight Path 12 (FP12) at LANSCE. Neutrons which are delivered to FP12 are produced by the 1L Lujan Target-Moderator-Reflector-Shield (TMRS) Mark-III assembly and are moderated by a partially-coupled cold hydrogen moderator. The flight path is equipped with an \(m=3\), 10 cm\(\times\)10 cm supermirror neutron guide designed to transport a large fraction of cold neutrons to the experimental hutch~\cite{Koehler2017}. The neutron flux and spectrum in the slow neutron regime between 1-80 meV was already measured in the past and described in the literature~\cite{Seo2004}. MCNPX simulations of FP12 were performed to extract the energy spectrum and emission time distributions of 0.7 eV neutrons. The detailed 3D geometry as implemented in the MCNPX model is shown in Figure \ref{figure:mcnp}. The model included a detailed implementation of the TMRS Mark-III surrounded by the Bulk Shielding. The neutron guide is surrounded with concrete/steel/polyethylene shielding and the experimental hutch is shielded with borated polyethylene clad in 3/8" thick steel. The neutron beam terminates in a beam stop made of steel and borated polyethylene.

The calculations of the time-emission spectra followed the methodology outlined in~\cite{Mocko2008}. The TMRS model used in these simulations was based on the as-built engineering design. The thermal neutron scattering kernels that were used are for hydrogen and deuterium in water, ortho- and para-hydrogen in liquid hydrogen, aluminum in alloys, and iron in steel. All calculations utilized the next-event-estimator (point detector) variance reduction technique~\cite{Pelowitz} to ensure efficient convergence of all extracted observables. The relatively sharp proton pulse (270 ns pulse width) incident on the spallation target at a short-pulsed spallation neutron source like LANSCE is a very valuable feature for precision measurements of parity violation and time reversal violation on neutron-nucleus resonances as it greatly improves the time resolution of the neutron time-of-flight measurements necessary to accurately determine the neutron energy.


\begin{figure}[H]
\centering
  \includegraphics[width=\linewidth]{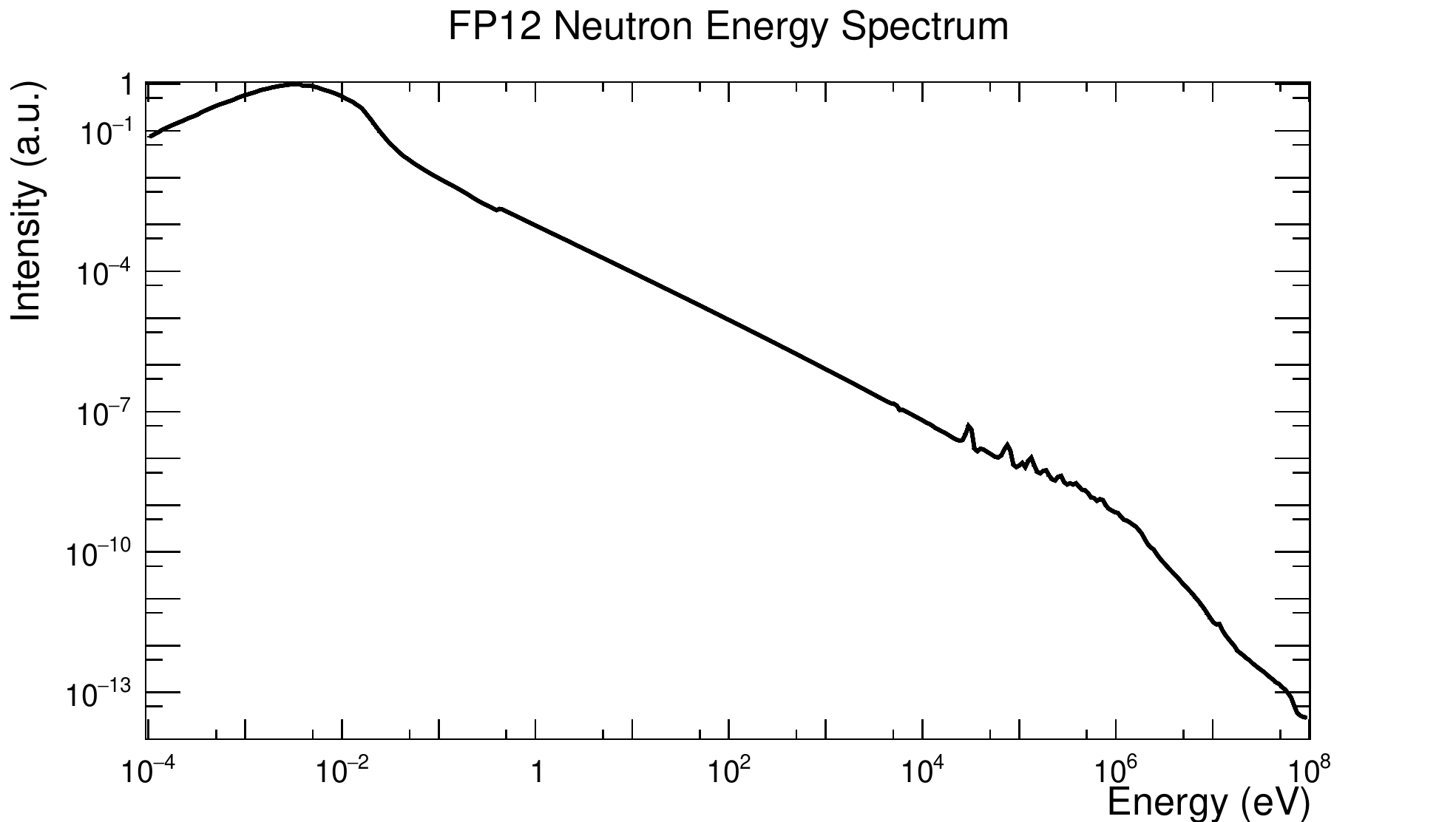}
\caption{An MCNPX calculation of the neutron energy spectrum for FP12, normalized to the peak flux intensity. This was generated using MCNPX simulations with the most recent FP12 geometry.}
\label{energyfp12}
\end{figure}

\begin{figure}[H]
\centering
  \includegraphics[width=\linewidth]{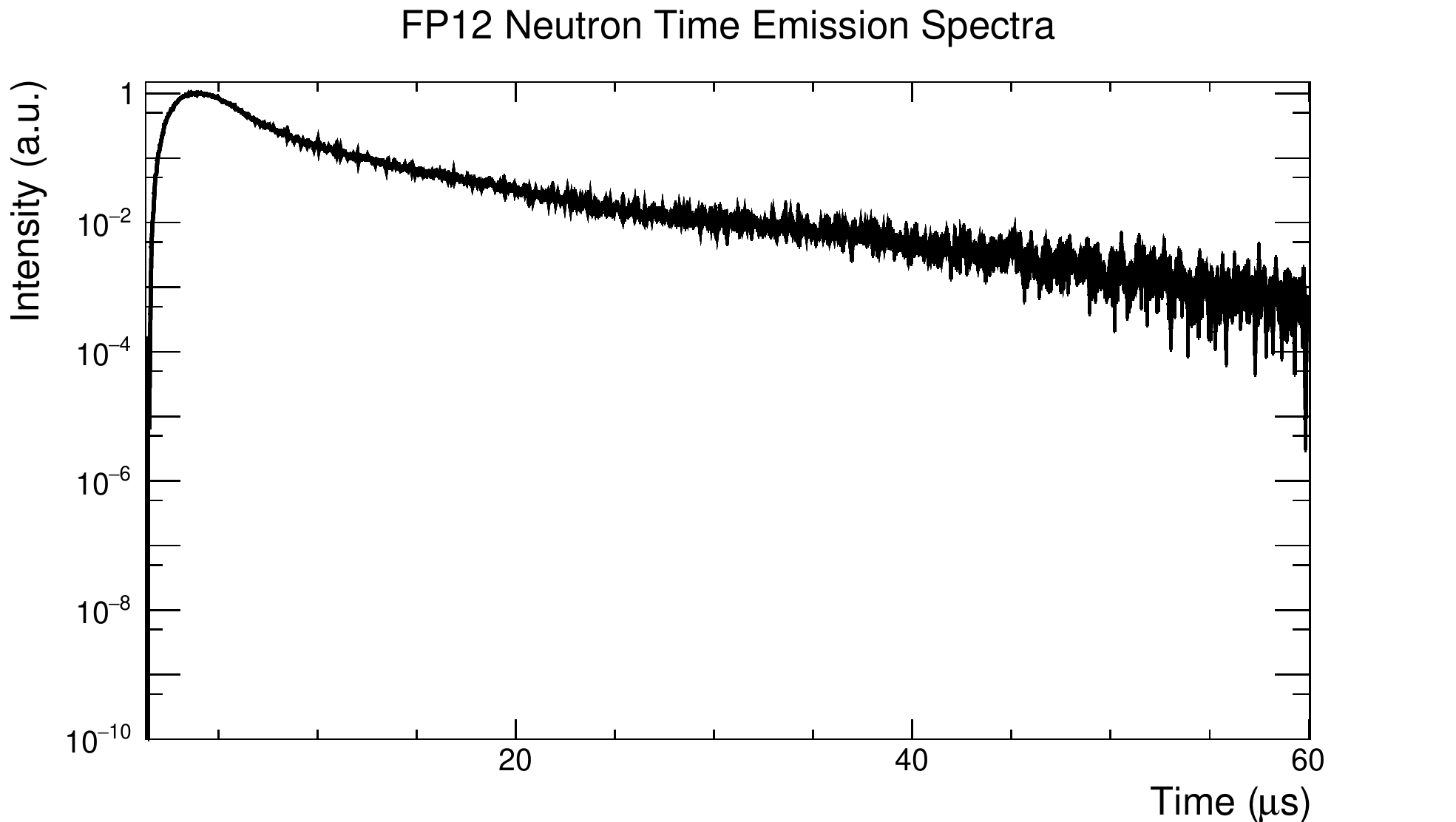}
\caption{A calculation of the time emission spectra for neutrons of energy 0.7~eV, normalized to the peak intensity.}
  \label{time07ev}
\end{figure}

\begin{figure}[H]
\centering
\includegraphics[width=\linewidth]{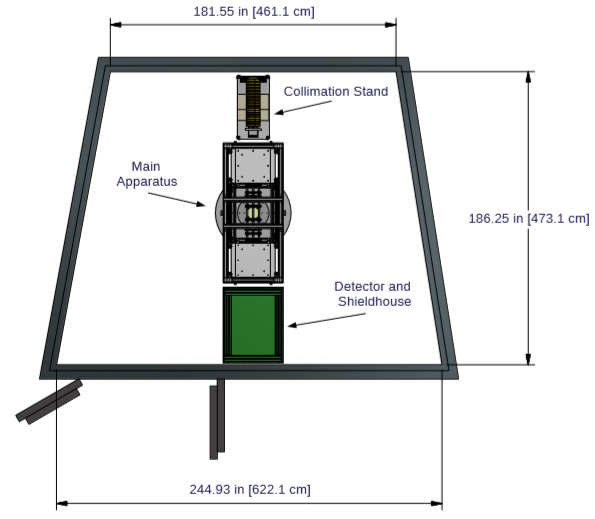}
\caption{Top-down view of the FP12 hutch at LANSCE showing the hutch dimensions and general position of the experimental setup.}
\label{fig:whole_trex_1} 
\end{figure}
\begin{figure}[H]
\centering
\includegraphics[width=0.9\linewidth]{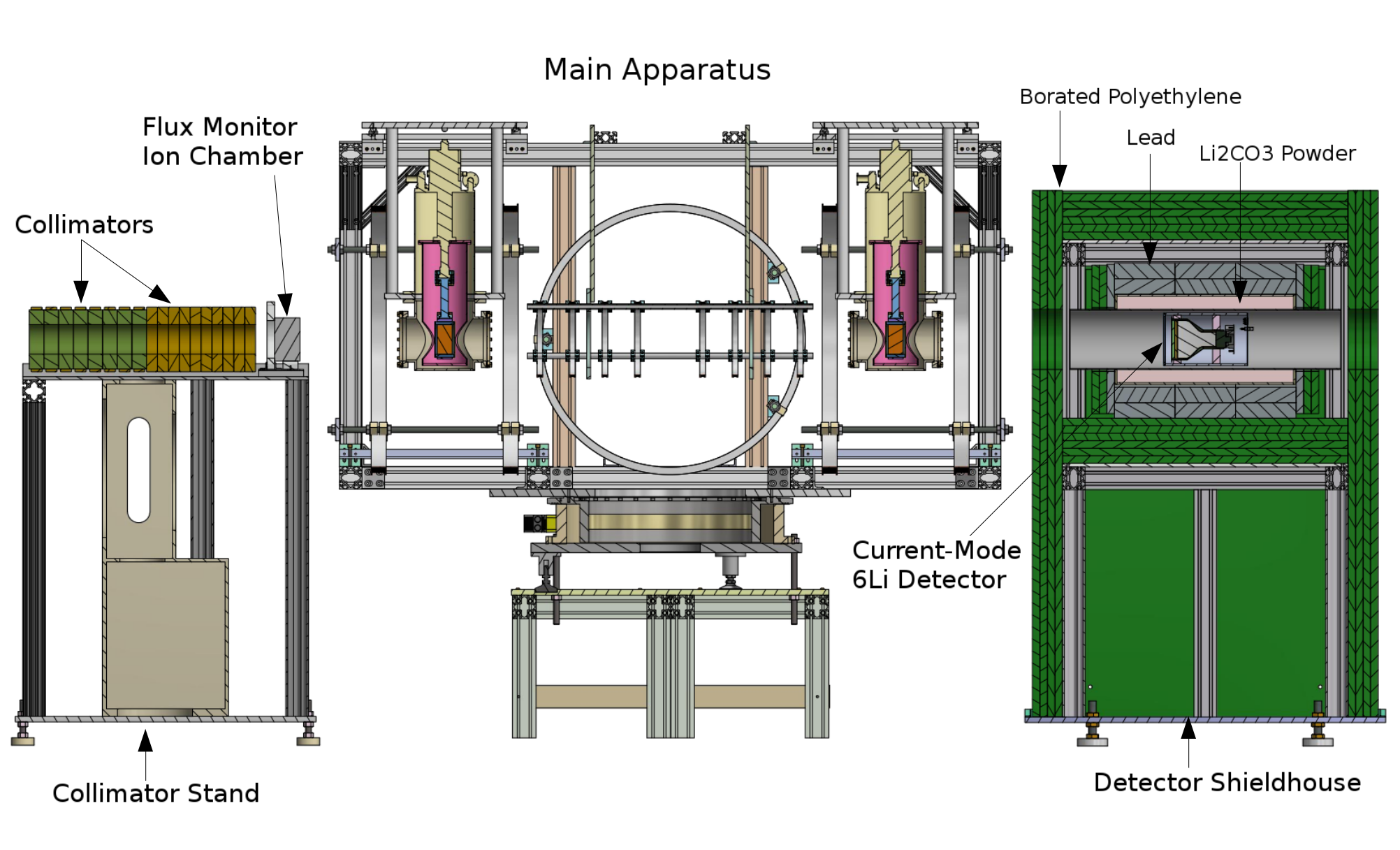}
\caption{ A cross-section view of the experimental setup in FP12 at LANSCE as viewed from the side. The three components are the collimator stand, the main apparatus containing the cryostats and the spin flipper coil assembly, and the detector/shieldhouse assembly.}
\label{fig:whole_trex_2}
\end{figure}

The FP12 hutch is a steel enclosure shaped like a trapezoid, as seen in Figure \ref{fig:whole_trex_1}. It is approximately 4.73 m long in the beam direction, 2.74 m tall, and 4.61 to 6.22 meters in the transverse direction. The beam pipe is centered 1.37 meters above the floor. The most upstream components are collimators that scrape the neutron beam to a 10 cm diameter and a $^{3}$He-$^{4}$He ion chamber to measure the neutron flux. Next downstream is the first cryogenically cooled $^{139}$La target, an adiabatic spin flipper, and a second $^{139}$La target. The furthest downstream component is a shielded $^6$Li-rich scintillator neutron detector buried in shielding designed to attenuate both neutrons and gamma rays to reduce background noise. Figure \ref{fig:whole_trex_2} shows a detailed view of these components and their placement.

\section{Mechanical Apparatus Components}\label{s:apparatus}

 Previous experiments have measured \(P\) on the 0.7 eV resonance in $^{139}$La to \(\sim4\%\) accuracy \cite{Yuan91}. In order to achieve a measurement with 1\% accuracy as needed for the future time reversal and neutron polarimetry physics, we decided to use cryogenic targets to minimize the Doppler broadening of the transmitted resonance profile. To prepare for the future NOPTREX time reversal search experiment described in \cite{Bowman2014}, we mounted these cryostats on a rotating turntable and chose a $^{3}$He neutron spin filter polarized by spin-exchange optical pumping (SEOP) methods which can later be placed in the position of the upstream cryostat. These decisions strongly influenced the apparatus design, which otherwise possesses several components very similar to an earlier instrument built by the TRIPLE Collaboration~\cite{Roberson93}. The apparatus also possesses some common features with the POLYANA instrument at the IBR-30 pulsed reactor neutron source at Dubna~\cite{Alfimenkov1991}. 
 
 The apparatus can be divided into three major beamline components; the collimator stand; the main apparatus stand (mounted on the rotation stage); and the detector stand/shieldhouse. One key feature of almost all components is the use of non-magnetic materials. Due to the oscillating magnetic fields produced by the neutron spin-flip coils, any magnetic material could be magnetized, leading to stray fields that could affect the neutron spin-flip efficiency as well as cause magnetic-field dependent gain shifts in our detector electronics. Therefore much of the apparatus is constructed from aluminum, brass, and plastic. 

\subsection{Collimators and Stand}
The collimator stand, located furthest upstream at the exit of the beam pipe, supports interchangeable collimators and the \(^3\)He-\(^4\)He beam monitor. The stand is made from aluminum held together with brass screws and bolts. A 96 cm long rectangular aluminum channel provides a reliable way to interchange up to 16 collimators and ensures that all collimators are coaxial with the beam. Adjustable feet allow for precision alignment of the collimators to the beam axis using a theodolite. The beam monitor rests at the downstream end of the stand and is electrically isolated from the rest of the stand via a polycarbonate plastic base. The collimators define the neutron beam and minimize the contribution of fast neutrons and gamma rays to the radiation backgrounds in the hutch. 5.01 cm thick borated polyethylene collimators attenuate fast neutrons outside of the desired beam profile. The borated polyethylene collimator located the furthest upstream has an additional 1~mm layer of \(^6\)Li-loaded fluorinated plastic which collimates slow neutrons and produces a very small number of gamma rays per absorbed neutron. After the borated polyethylene collimators, 3.65 cm thick brass collimators attenuate gamma rays produced both by the `gamma flash' characteristic of spallation targets as well as any gamma rays produced by neutrons absorbing in the borated polyethylene collimators. The collimator collection is also draped with $^{6}$Li fluorinated plastic to absorb high energy neutrons scattered by the hydrogen but not absorbed by the boron in the fast neutron collimators before they can create additional neutron and gamma ray backgrounds signals in the detector. Collimators made of $^{10}$B-loaded plastic are also placed outside the upstream ends of the cryostat vacuum flanges to sharply define the beam that passes through the targets.  The stand is stable enough to also support a calibration resonance target wheel and a local neutron chopper if needed. 

We measure the neutron time-of-flight spectrum on a pulse-by-pulse basis using a beam monitor, which is a low efficiency ion chamber with $^{3}$He and $^{4}$He gas chambers located back-to-back along the neutron beam and operated in current mode. Since the neutron absorption cross section in $^{3}$He is very large (of order kilobarns) and well known~\cite{Keith2004}, the neutron absorption cross section in $^{4}$He is extremely small, and the gamma interactions of these two isotopes are essentially identical, the difference signal from these two chambers is directly proportional to the instantaneous neutron flux. We do not elaborate any more on the details of the design for this device as it is identical to the one used by the TRIPLE collaboration as it is already described in great detail in the scientific literature~\cite{Szymanski}.   


\subsection{Main Apparatus and Rotation Stage}

\begin{figure}[H]
\centering
\includegraphics[scale=0.45]{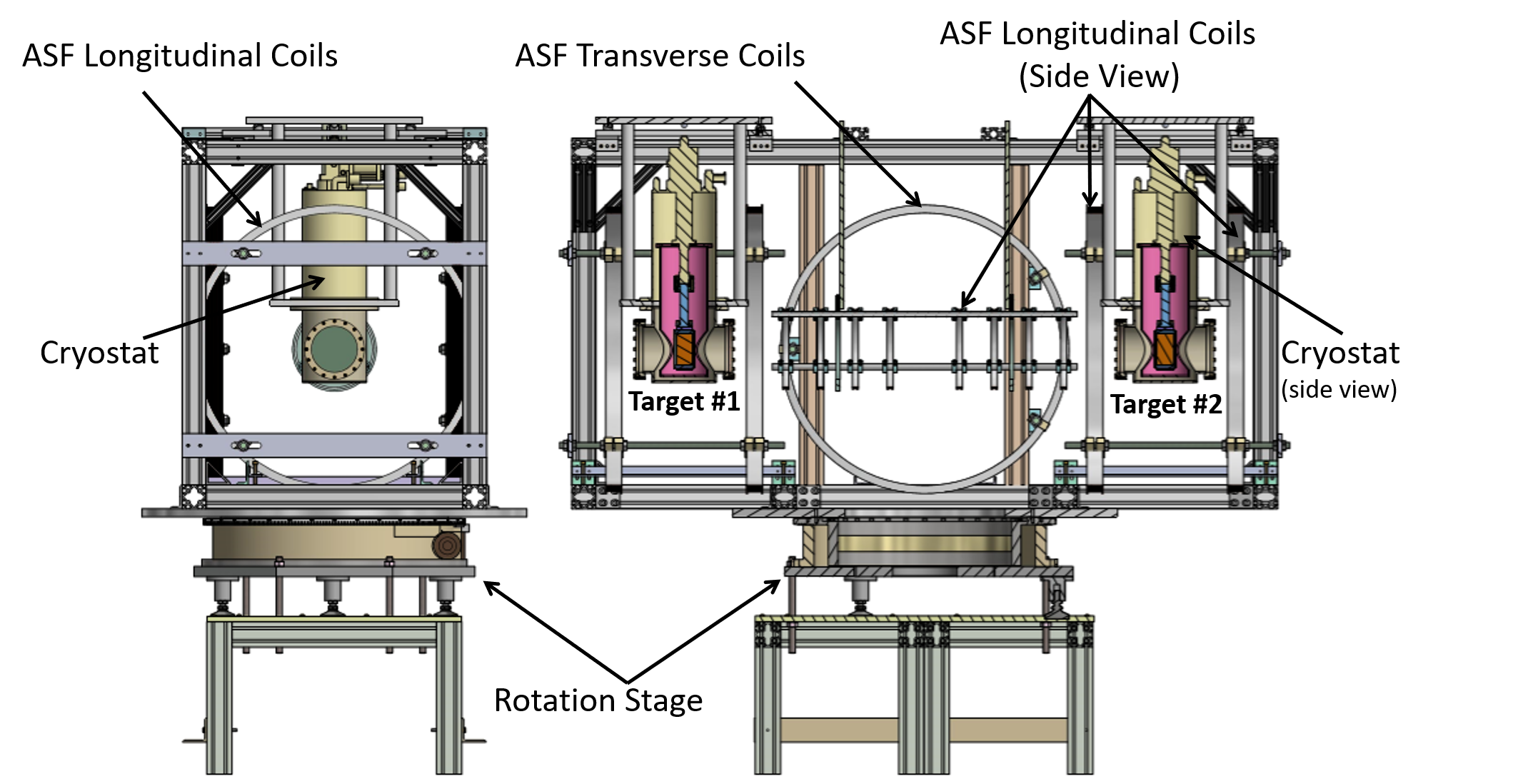}
\caption{The main apparatus in the Double Lanthanum experimental setup as configured with two cryogenic $^{139}$La targets installed. The lefthand diagram shows a downstream view of the apparatus (i.e. looking down the axis of the neutron trajectory) whereas the righthand shows a sidelong, cross-section view.}
\label{fig:main_apparatus}
\end{figure}

The main apparatus frame is constructed from extruded aluminum profiles (manufactured by 80/20 Inc, hereafter referred to as ``8020") mounted to a large rotation stage with adjustable feet. The base of the main apparatus is a large aluminum table with large 8020 support legs. On top of this table sits a Franke TSD-830M rotation stage, labeled in Figure \ref{fig:main_apparatus}. The rotation stage has a 57,000~N load rating and allows for 360 degree rotation of the main apparatus. The rotation stage is mounted to adjustable feet which allow for fine-tuned adjustments to center the targets on the beam axis. A large (1.22~m diameter and 2.54 cm thick) circular aluminum plate rests on the rotation stage and supports the main rectangular frame constructed from 8020. The frame supports the cryostat/cryogenic target assemblies, adiabatic spin-flipper coils, and can be modified to include additional equipment such as a \(^3\)He polarizer (described in further detail in Section \ref{s:3He}). The spin-flipper coils are affixed directly to the 8020 aluminum frame using specially-designed plastic clamps. 

The aluminum housing for the cryostats are supported from the top of the 8020 frame on sliding rails, allowing for easy adjustment of the target location along the neutron beam. The maximum distance between the two targets is 152 cm. The cryogenic housing is mounted to the sliding rails by three V-groove-and-ball kinematic mounts. This allows for reproducible, precision placement of the targets when moving or replacing the cryogenic housing during a target change. In this modular design other components such as a large \(^3\)He polarizer can also couple to the sliding rails or the 8020 frame.  One nice feature of using 8020 for the frame is the ease with which components can be precisely and repeatably aligned using mechanically-defined reference points. The aluminum table was aligned to the beam using a theodolite. All other components used mechanically-defined reference points to define alignment.

\subsection{Target Environment}
Cryo-Torr 8 cryopumps were stripped of their internal cryopumping surfaces and used as mechanical refrigerators to cool the targets to \(\sim\)15 K.
The cryopumps couple to the aluminum housing and then to an aluminum vacuum chamber below. The neutron beam passes through 1.27 mm thick aluminum vacuum windows centered on the beam. Inside the cryostat vacuum chamber, the target cell is thermally coupled to the 15 K stage of the cold head, while a surrounding radiation shield is coupled to the 80 K stage of the cold head. The target cells are contained in 12.03~cm diameter aluminum cans, filled in an inert argon atmosphere in a glovebox and sealed with indium o-rings. This prevents potential oxidation of sensitive or reactive target materials such as lanthanum to preserve the integrity of the target and to promote the safe handling of activated targets. To monitor the temperatures inside of the cryostats, four silicon diode temperature sensors were placed at different stages of the cryostat; one on the radiation shield, one on the coldhead, one at the top of the cell clamp, and one at the bottom of the cell clamp. The temperatures measured by diode thermometers were read using a Lakeshore 218 temperature monitor. The temperature distribution in the cryogenic system under steady state operation was stable at the level of 0.2 K. 


\subsection{Detector and Shieldhouse}
Located furthest downstream from the beam pipe entrance to the hutch is the detector stand and shielding assembly, known as the shieldhouse. The purpose of this shieldhouse is to protect the  \(^6\)Li neutron detector from any neutrons and gamma rays  originating from outside the defined neutron beam (e.g. multiply-scattered neutrons from the hutch or gamma rays produced in neutron capture reactions on materials in the hutch). The support structure frame is made of 8020 and sheets of borated polyethylene. The outermost layer of shielding consists of 15~cm of borated polyethylene, followed by 10~cm of lead bricks and 5~cm of lithium carbonate powder as seen in Figure \ref{fig:whole_trex_2}. The detector sits in an aluminum tube and was aligned to the neutron beam axis using a theodolite and crosshairs of fishing line anchored to the aluminum tube using offset set screws.


\subsection{Field Mapper}
\label{s:mapper}

In order to successfully make a precision measurement of the parity violation in the target nuclei, it is critical to understand the dynamics of the neutron spin motion in the fields and determine the spin-flip efficiency. To do so, detailed maps of the magnetic fields produced by the spin flipper array (discussed in more detail in Section \ref{s:spinflipper}) were needed.

One of the challenges presented with the design of the spin flipper coils was a difficulty in mapping the magnetic field. Mounting a magnetic probe to a motorized mapper system was very difficult due to spatial constraints enforced by the coils, their supports, and lead wires. A simple mapping system was constructed by mounting a Lakeshore 460 triple axis probe to a three-dimensional, manually translatable apparatus. Continuous translational range of motion was possible in the \(xy\) plane perpendicular to the direction of the neutron beam. In the longitudinal $z$ direction, pairs of threaded holes were machined into a long, metal plank; pairs of screws were then placed in each set of holes and by laying the y-directional translation stage such that an edge was laid flush to both screws, unique points along the \(z\) axis were defined. Cross-sectional \(x,y\) slices of the magnetic field were then mapped for each of the points defined along the $z$-axis. 12 planar maps were taken for each configuration of the spin flip field. This data, shown in Figure \ref{fig:SF Field Maps}, was then reconstructed and interpolated using COMSOL and used in calculating the neutron spin flip efficiency as discussed further in Section \ref{s:spinflipper}.

\begin{figure}[H]
\centering
\includegraphics[width=0.9\linewidth]{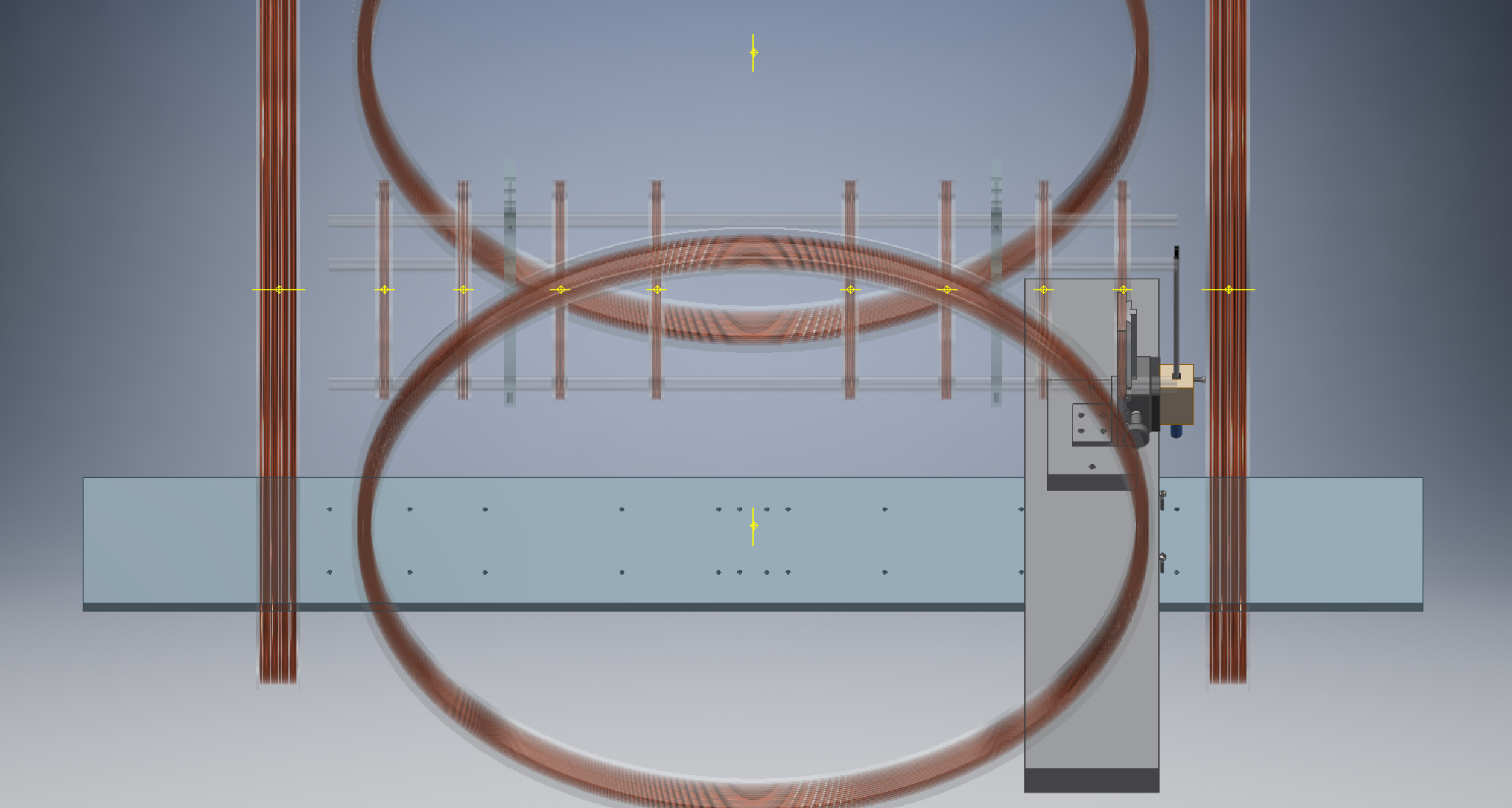}
\caption{\label{fig:mapper} CAD drawing of the mapper used to map the B-field. Note the pairs of holes machined into the longitudinal plank; these were used to uniquely define points along the $z$-axis for maps to be taken.}
\end{figure}

\section{Adiabatic Spin Flipper for eV Neutrons}\label{s:spinflipper}
\label{S:4}

\begin{figure}[H]
\centering
  \includegraphics[width=\linewidth]{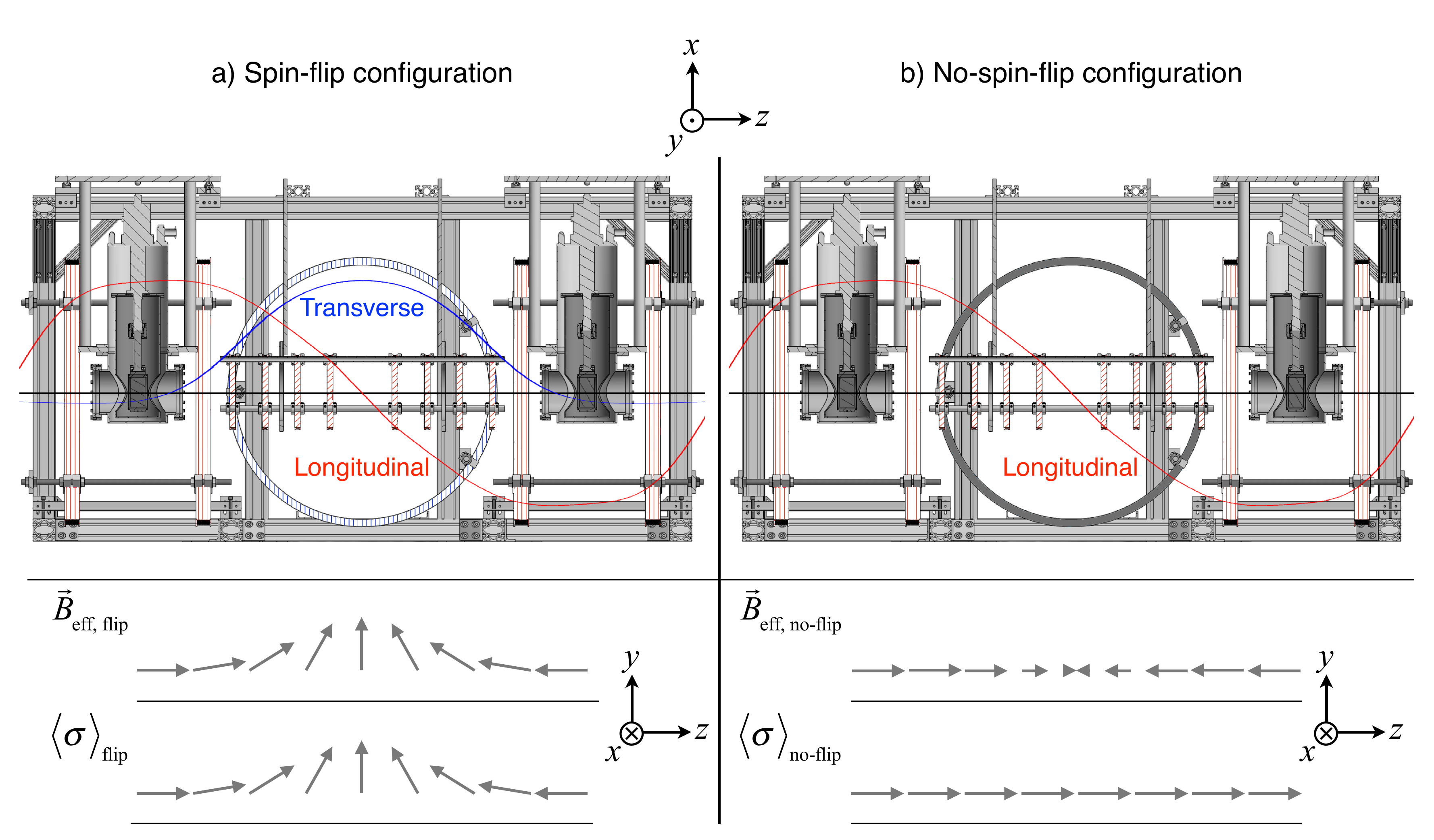}
 \caption{Components of the magnetic field in the a) spin-flip and b) no-spin-flip configurations. The elements of the apparatus producing each component of the magnetic field (longitudinal -red- coils and transverse -blue- coils) are highlighted. Also the total magnetic field $\vec{B}_{\textnormal{total}}$ and the projection of the neutron spin $\langle \vec{\sigma} \rangle$ along its axis of precession are shown in the $zy$ plane for each case.}
 \label{fig:field_apparatus}
\end{figure}

Following the design of Roberson et al. \cite{Roberson93}, we simulated and built a neutron Adiabatic Spin Flipper (ASF) based on adiabatic spin motion in static magnetic fields. This spin flipper is mounted and aligned to the 8020 frame, ensuring that the ASF longitudinal coils shown in Figure \ref{fig:main_apparatus} are coaxial with both the lanthanum targets and the neutron beam. The coil geometry and currents were optimized for neutron energies near 1~eV as needed for the Double Lanthanum experiment, though the adiabatic condition can be met for a broad range of neutron energies. As in the TRIPLE apparatus, the neutron spin flipper consists of longitudinal and transverse coils that produce static gradient magnetic fields. The combination of these fields produces a total field with approximately constant amplitude \(B_0\) that turns the neutron spin over the distance \(L\) on the \(z\)-axis over which the spin rotation occurs. The spin flipper operates in two field configurations, which we shall call the \textit{no-flip} and \textit{flip} configurations. For the \textit{no-flip} configuration, only the longitudinal \(B_z\) coils are turned on. If the direction of the neutron beam is along the \(z\)-axis and \(x=y=z=0\) is taken to be the center of the spin flipper as shown in Figure \ref{fig:field_apparatus}, then the magnetic field produced by the longitudinal coils in the \textit{no-flip} state can be approximated as
\begin{equation}
\label{eq:sin-cos1}
  B_z =
  \begin{cases}
                                   B_0 & \text{if \(z<-L/2\)} \\
                                   -B_0\sin{(\pi z/L)} & \text{if \(-L/2 \leq z\leq L/2\)} \\
  -B_0 & \text{if \(z>L/2\)}
  \end{cases}
\end{equation}
Recall that the magnetic torque \(\vec{\tau}\) experienced by a particle with a magnetic moment \(\Vec{\mu}\) in an external magnetic field \(\vec{B}\) is 
\begin{equation}
\Vec{\tau}=\Vec{\mu}\times\vec{B}
\end{equation}


 For the \textit{flip} configuration, the transverse $B_y$ coils are turned on (in addition to the $B_z$ coils). The component produced by the transverse coils is given by

 \begin{equation}
\label{eq:sin-cos2}
B_{y}=\left\{
	\begin{array}{ll}
		\pm B_{0}\cos(\pi z/L)  & \mbox{if } -L/2\leq z \leq L/2\\
		\;\;\;0 & \mbox{otherwise} 
	\end{array}
\right.
\end{equation}


The $\pm$ sign indicates that the $B_y$ component can be either parallel or anti-parallel to the $y$-axis, depending on the direction of the current in the transverse coils. The superposition of the $B_z$ and $B_y$ fields produces a field $\vec{B}_{\textnormal{total}}$ that is constant in magnitude but rotates by $180^{\circ}$ in the $zy$ plane over the length $L$. A positive $B_y$ component will produce a counter-clockwise rotation of the total field in the $zy$ plane, while a negative $B_y$ component produces a clockwise rotation.

As the neutron magnetic moment $\vec{\mu}_n$ is related to the neutron spin $\vec{\sigma}$ by the gyromagnetic ratio $\gamma_n$, one can construct the Hamiltonian of the magnetic interaction with an external magnetic field $\vec{B}$. Substituting this Hamiltonian into the Schr{\"o}dinger equation for a time dependent $\vec{B}$ field, the Larmor equation for the spin can be obtained \cite{Kraan2004}: 

\begin{align}
  \frac{d}{dt}\vec{\sigma}(t)=\gamma_n\vec{\sigma}(t)\times{}\vec{B}(t)
  \end{align}

The change in the neutron spin direction is normal to both $\vec{B}$ and $\vec{\sigma}$ at any given time. For the \textit{no-flip} state there should not be a change in the neutron spin as the spin and the magnetic field are parallel. If the field as seen in the rest frame of the neutron varies slowly enough (i.e. adiabatically), the neutron spin will precess about the field direction with the Larmor frequency $\omega_{L}=\gamma_{n}B$ and will ``follow'' the direction of the external field from its initial direction $+z$ to the opposite direction $-z$ at the end of the flipper. The rotation frequency $\omega_B$ of the magnetic field as seen in the rest frame of the neutron is given by

 \begin{align}
     \omega_B=\frac{\pi}{t}=\frac{\pi}{L/v}
 \end{align}
 where $t$ is the time the neutron spends in the spin flipper and is given by $L/v$, where $v$ is the neutron speed and $L$ is the length of the spin flip region. The ratio of the field rotation frequency \(\omega_B\) and the Larmor frequency \(\omega_L\), is defined as the adiabaticity parameter for this system:
 \begin{align}
     \gamma=\frac{\omega_B}{\omega_L}=\frac{\pi v}{\gamma_n LB}
 \end{align}
 
 If $\gamma<<1$ the spin direction will undergo several rotations around the $\vec{B}$  direction for every small variation of the field, keeping its precession axis approximately aligned with $\vec{B}$ at any given time. In other words, as long as the neutron speed $v$ is below a certain limit, the transport of the spin will be adiabatic.
  
  Figure \ref{fig:field_apparatus} shows the magnetic field configuration for the \textit{flip} and \textit{no-flip} states of the spin flipper, as well as the total magnetic field and neutron spin projection along the field direction as neutrons propagate across the spin flipper length.
  

This simplified description assumes that all of the neutrons travel directly down the longitudinal axis of the coils. Realistic spin flipper efficiency calculations must account for the finite radius of the beam as well as deviations away from the on-axis magnetic field. Bowman, Penttil\"{a}, and Tippens~\cite{Tippens95} investigated these effects for neutrons whose trajectories deviate slightly from the axis of symmetry. One must consider not only the efficiency of the spin flipper in the \textit{flip} configuration but the efficiency in the \textit{no-flip} configuration. The total efficiency  of the spin flipper is the average of the efficiencies for the \textit{flip} state and the \textit{no-flip} state:

\begin{equation}
\label{eq:flip}
\epsilon_{\textnormal{flip}}=1-\frac{1}{\sqrt{1+\gamma^{-2}}}\left[1-\cos\left(\pi\sqrt{\gamma^{-2}+1}\right)\right],
\end{equation}

and the spin-preserving efficiency in the \textit{no-flip} state is
 
 \begin{equation}
  \label{eq:no-flip}
\epsilon_{\textnormal{no-flip}}=1-\frac{\pi^{3}r^{2}}{8\gamma L^{2}},
\end{equation}

with $r$ the distance of the neutron trajectory from the spin flipper axis. 
It can be seen that $\epsilon_{\textnormal{flip}} \rightarrow 1$ when $\gamma\rightarrow0$; the smaller the adiabaticity parameter, the higher the spin-flip efficiency. Also, $\epsilon_{\textnormal{no-flip}}=1$ when $r=0$; if all neutrons travel over the beam axis every spin remains unchanged in the \textit{no-flip} state.

It now makes sense to define a `total' efficiency, \(\epsilon_{tot}\), as a figure of merit for the design which describes both how well the \textit{flip} configuration flips the spin and how well the \textit{no-flip} configuration preserves the spin. We therefore define 
\begin{equation}
    \label{eq:total}
     \epsilon_{\textnormal{total}}=\frac{\epsilon_{\textnormal{flip}}+\epsilon_{\textnormal{no-flip}}}{2}
 \end{equation}
 For an ideal spin flipper that flips all neutron spins in the \textit{flip} configuration and leaves all neutron spins unchanged in the \textit{no-flip} configuration, we see \(\epsilon_{tot}=1\).

\begin{figure}[H]
\centering
\includegraphics[width=\linewidth]{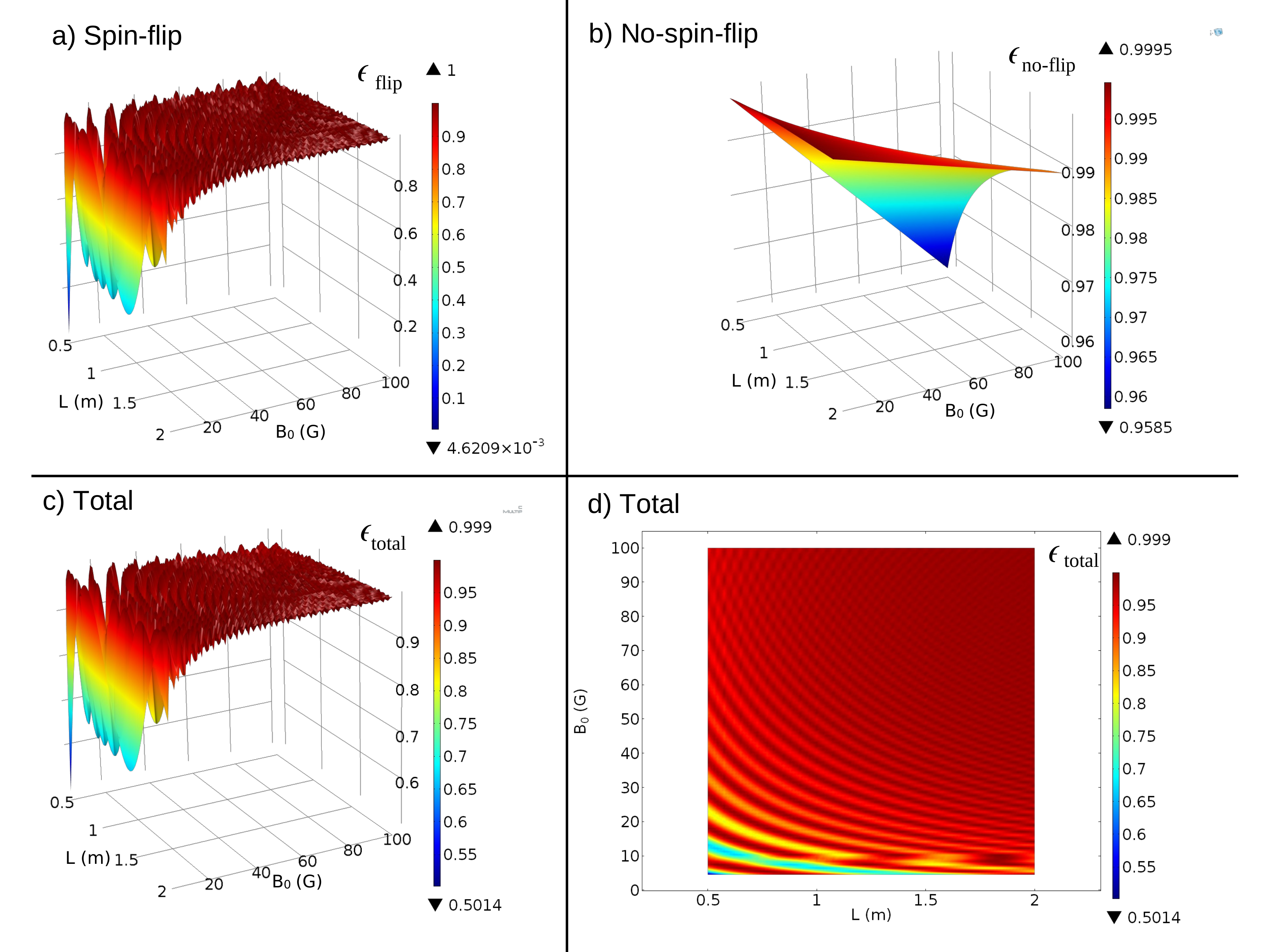}
        \caption{Calculations of the a) spin flipper efficiency and b) spin preserving efficiency as functions of the parameters \(B_0\) and \(L\) for 1~eV neutrons at \(r=3.54\)~cm, the value of \(\sqrt{\langle r^2\rangle}\) over a 10 cm diameter beam. The total efficiency is shown in c) and d).}
        \label{fig:efficiency}
\end{figure}
\begin{figure}
    \centering
    \includegraphics[width=\textwidth]{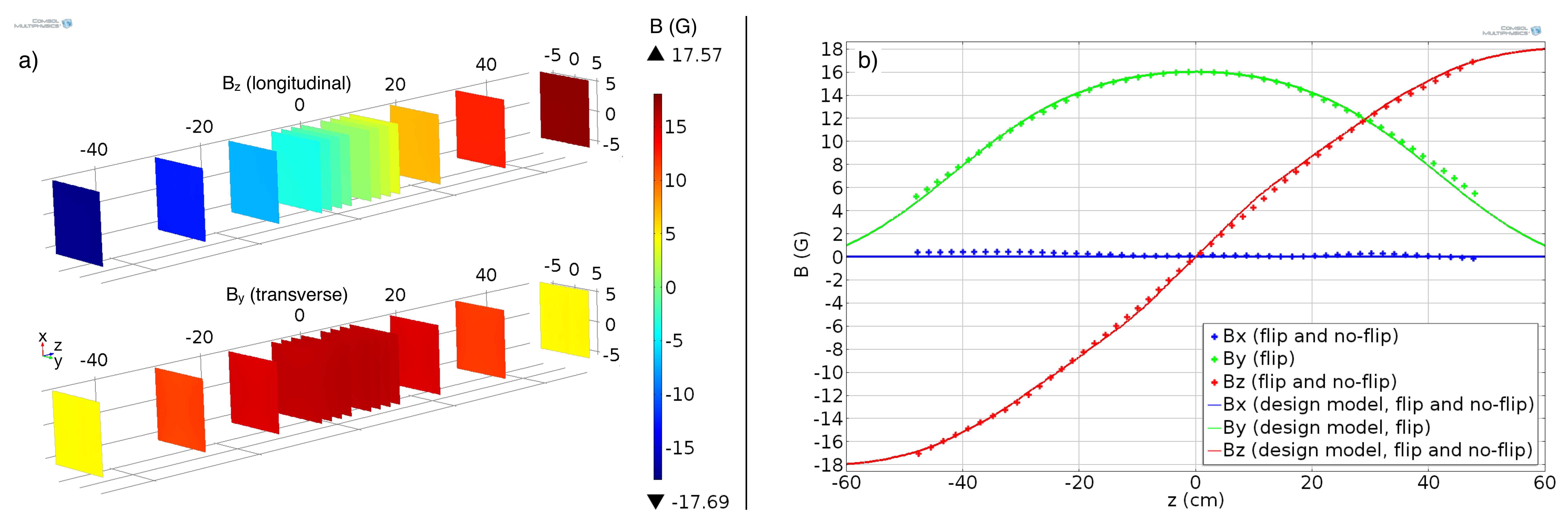}
    \caption{a) Field maps for the longitudinal and transverse components of the spin flipper; 14 $xy$ planes were scanned at positions between -50 to 50 cm along the $z$-axis. b) Measured magnetic field components along the $z$-axis and their comparison with the design model fields.}
    \label{fig:SF Field Maps}
\end{figure}

To define the optimum parameters for our spin flipper, we calculated the \textit{flip, no-flip} and total efficiency for different values of magnetic field amplitude $B_0$ and spin flipper length $L$ using equations \ref{eq:flip}-\ref{eq:total}. We assumed a 10 cm diameter neutron beam and used the square root of the average value of $r^2$, $\sqrt{\langle r^{2} \rangle}$=3.54 cm to calculate the spin-preserving efficiency. Figure \ref{fig:efficiency} shows these calculations. Considering the constraints in length for the Double Lanthanum experiment at FP12, we chose $L=120$ cm for the flipping length, the maximum allowed in the apparatus. From $z=-60$ cm to $z=60$ cm the field $\vec{B}$ has the described form for the \textit{flip} and \textit{no-flip} states (equations \ref{eq:sin-cos1} and \ref{eq:sin-cos2}). For 60 cm $<\lvert z\rvert<$ 75 cm we aimed to produce a longitudinal constant field of amplitude $\pm B_0$, providing a uniform field in the region of the La targets (see Figure \ref{fig:SF Field Maps}). The choice of $B_0$ values were constrained by practical experimental and safety reasons: it has to be generated by reasonably attainable electric currents and voltages, yet also has to be large enough to assure a good efficiency. We found that for $L=120$~cm, a magnetic field amplitude of $B_0$=16-17 G can produce a total efficiency between 92\%-98\%, as seen in Figure \ref{fig:efficiency}. 

To establish the position along $z$ and the parameter $NI$ (number of turns of a particular coil multiplied by the current) for each of the coils comprising the longitudinal field array (see Figure \ref{fig:SF Field Maps}), we considered the current $I$ to be fixed, and in an iterative process of varying the position and obtaining the number of turns by the Single Value Decomposition (SVD) method, we obtained the optimum coil configuration. With $I=15$ A and a maximum number of turns that vary from 1-2 turns for the small coils to $\sim$80 for the large coils, the voltage requirement using 10 AWG copper wire is close to 40 V, which is achievable with the system described in Section \ref{s:apparatus}. 

The field maps of the actual spin flipper that was constructed on FP12 were obtained using the mapper described in Section \ref{s:mapper}. A total of 14 $xy$ planes of 11 cm $\times$ 11 cm of cross section around the SF axis were scanned in steps of $5$ mm. The field maps span the region between -50 to 50 cm in the $z$-axis, as shown in Figure \ref{fig:SF Field Maps}a. A comparison of the measured magnetic fields and the initially calculated magnetic fields along the spin flipper axis is shown in figure \ref{fig:SF Field Maps}b; a good agreement is observed in general, however it is important to point out that the longitudinal magnetic field (pictured in red), has a higher amplitude than the transverse magnetic field, producing a total magnetic field that, although performing the desired rotation by 180$^\circ$ in the $yz$ plane, does not maintain a constant amplitude. The field map was obtained in the middle of the experimental run; therefore the spin flipper was operated with the magnetic fields configuration shown in figure \ref{fig:SF Field Maps}. 

These field maps, in combination with the images of the neutron beam profile and data from a Monte Carlo neutron spin transport simulation, can be used to estimate the actual spin flipper efficiency 0.7 eV neutrons.

\subsection{Neutron Beam Intensity Maps}

\begin{figure}[H]
\centering
\begin{subfigure}[t]{.5\textwidth}
  \includegraphics[scale=0.5,left]{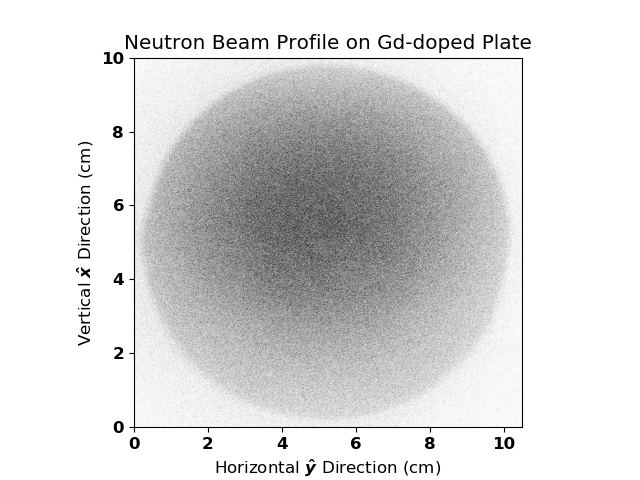}
  \subcaption{}
 \label{fig:intensity1}
\end{subfigure}%
\begin{subfigure}[t]{.5\textwidth}
  \includegraphics[scale=0.5]{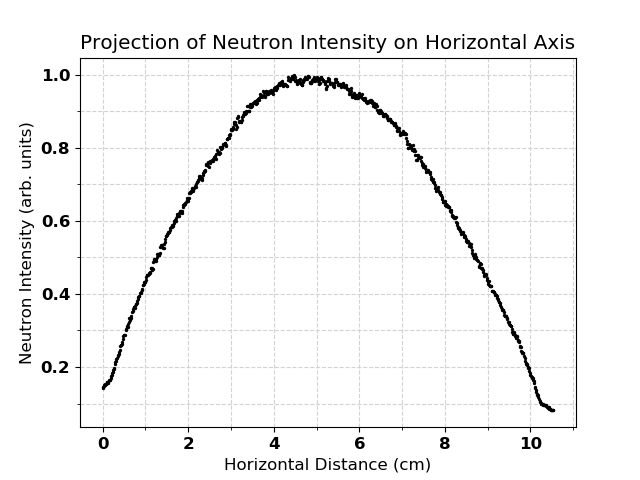}
  \subcaption{}
  \label{intensity2}
\end{subfigure}
\caption{The image on the left shows the neutron beam spot as imaged on a Gd-doped image plate. The neutron beam intensity distribution was extracted using ImageJ software.}
\end{figure}

The radial dependence of the neutron beam intensity must be mapped to sufficient accuracy that it can be included with the magnetic field map to calculate the neutron spin flipper efficiency. We determined that an intensity map with a few mm spatial resolution over the 10 cm diameter beam would suffice to determine the spin flip efficiency with an accuracy more than one order of magnitude better than our statistical accuracy goal. This measurement was performed with a commercial imaging plate using a neutron-sensitive Gd-doped film. Its spatial resolution is one order of magnitude better than required for our purposes. Figure \ref{fig:intensity1} shows an example of such a map. The neutron imaging plate technology used to produce these intensity maps has been demonstrated to possess a linear response over a dynamic range of about 4 orders of magnitude as determined in the course of careful studies conducted at NIST for a Penning-trap-based neutron lifetime experiment~\cite{Nico2005}.

\section{Mechanical Design of the Adiabatic Spin Flipper and Controls}\label{s:electronics}
The spin flipper consists of twelve axially concentric longitudinal coils of different radii and winding numbers connected in series to produce the sinusoidally varying longitudinal field along the \(z\)-axis (Eq. \ref{eq:sin-cos1}) and one pair of axially concentric Helmholtz coils that produce the cosinusoidally varying transverse field along the \(y\)-axis (Eq. \ref{eq:sin-cos2}). The coils were designed using COMSOL to determine the coil parameters (coil dimensions, number of turns, positions, and necessary currents) needed to produce the desired fields. In addition, one pair of `shunt' coils was constructed, consisting of two coils identical in construction to the transverse field coils. Because of the relatively large (\(\sim15\)~A) currents needed to produce these fields and the relatively short timescales (\(\sim\)100~ms) needed for the transverse field on/off transitions, the shunt coils were introduced to the circuit so that instead of turning the current in the transverse coils on and off and potentially producing electronic crosstalk, the current was instead diverted to the shunt coils via a switch box module.  The longitudinal and transverse coils were mounted to the frame of the apparatus and the shunt coils were placed in a corner of the experimental hutch as far away as possible from the spin-flip region of interest, approximately 3 meters from the center of the spin flipper. The field produced by the shunt coils at the center of the spin flipper was measured and determined to be negligible.

\subsection{Switch Box Design}
\label{subsection:switchbox}

To control the neutron spin flip coils, we designed a switch box containing an array of FET switches with opto-isolated gate drivers controlled by TTL signals sent from an Arduino Mega 2560 board. The switch box was designed to be able to handle a maximum current of 20~A and to react quickly enough during a change of state that viable neutron pulses were not lost due to a slow slew time for the switch box. The limiting factor for the switching rate was found to be the intrinsic 120~ms settling time that it took for the currents to achieve a steady state in the coils. This characteristic time was due to the large inductance of the coils and the settling time of the eddy currents in the aluminum coil frames.

The switch box allowed complete control over the state of the spin flipper. By sending the appropriate logic signal from the Arduino board to the switch box, current to the coils could be turned on/off, reversed in polarity, or diverted to/from the shunt coils. The Arduino board was connected to the external Lujan Center \(t_0\) signal as a trigger source, and its output 5 V logic signals were sent through a simple conditioning circuit before entering the switch box. Three LEDs were included in the conditioning circuit to visually confirm that the state of the Arduino and that the states were changing as expected.


\begin{figure}[H]
\centering
  \includegraphics[width=\linewidth]{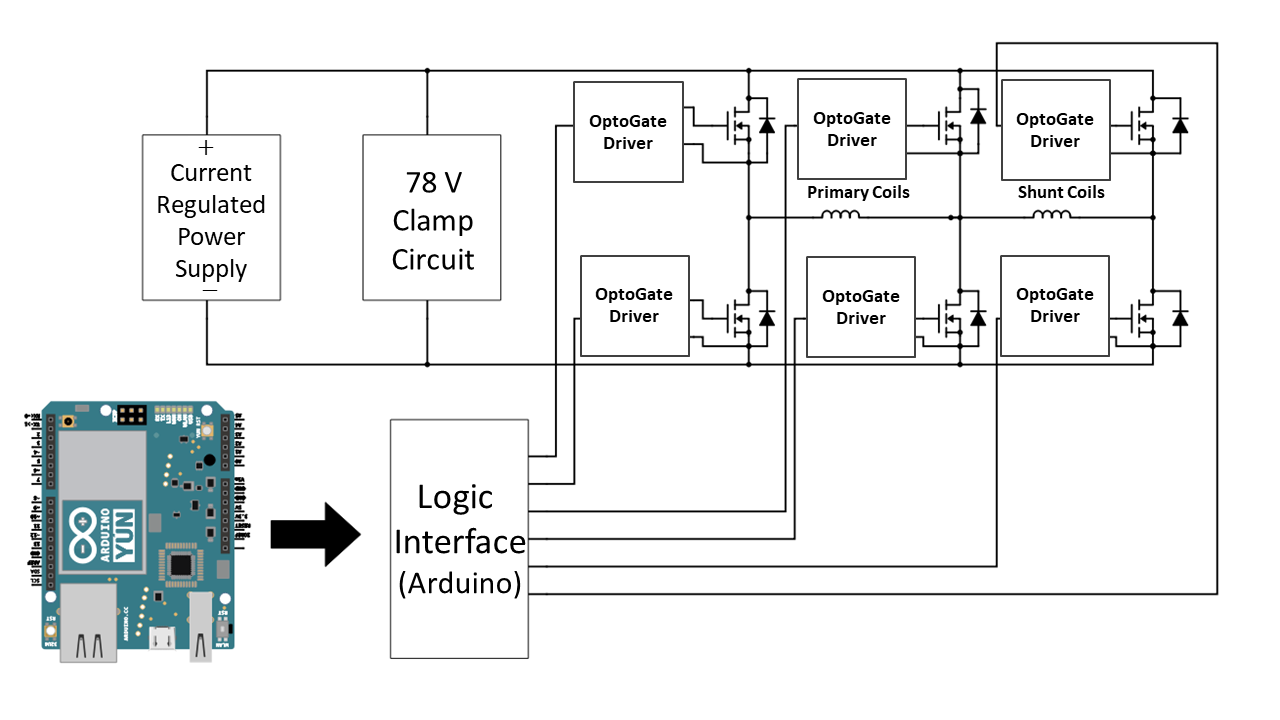}
 \caption{A simplified schematic of the spin flipper circuit including the primary transverse coils and the shunt transverse coils. The circuit was controlled by an Arduino Mega 2560 board.}
 \label{fig:switchbox}
\end{figure}

\subsection{Neutron Spin Flip Sequencing}

The choice of neutron spin flip sequencing is a critical factor in avoiding systematic errors that would create a false asymmetry and compromise the validity of this experiment. For example, the gain of the detector may drift linearly in time due to effects such as temperature-dependence of the detector electronics. Even more dangerous are gain effects that correlate directly to the absolute spin state of the neutron; stray magnetic fields produced by the spin flipper may influence the gain of the PMT dynode chain and cause the detector to have a different gain for each neutron spin state. An important feature of the spin flipper is that it works equally well for both configurations (\(\pm\)) of the field produced by the transverse coils. Therefore, a careful choice of the pattern with which the neutron spins are flipped can cancel systematic effects. Following the prescription by Roberson et al \cite{Roberson93}, the base spin sequence pattern F N N F N F F N was chosen, where N and F denote the \textit{no-flip} and \textit{flip} states, respectively. A discussion of how the spin state was determined during the data analysis process is discussed in section \ref{subsec:daq_faraday}.

Because the neutron spin flip behavior is identical for the cases where the transverse field is aligned (\(+\)) or anti-aligned (\(-\)) with the \(y\)-axis and recalling that the transverse field off (0) is the \textit{no-flip} state, we can additionally alternate the polarity of the field used to produce the \textit{flip} states, making our spin flip sequence + 0 0 - 0 - + 0. One can write the change in the neutron yield as a function of the stray field produced by the spin flipper, \(B_s\), as \begin{equation}
Y(B_s)=Y(B_0)+\frac{\partial Y}{\partial B}B_s+\frac{1}{2}\frac{\partial^2 Y}{\partial B^2}B^2_s+...
\end{equation}

then the difference in the neutron yields for the \textit{flip} and \textit{no-flip} states, \(\Delta Y\), is found to be \cite{Tippens95}: 

\begin{equation}
    \Delta Y = \overline{Y(F)}-\overline{Y(N)}=\frac{1}{2}\frac{\partial^2Y}{\partial B^2}B^2_s+...
\end{equation}
We can see here that the dependence on a term linear in \(\vec{B_s}\) vanishes, leaving us to only have to measure the change in the gain of the magnetic field at the detector location to estimate the quadratic term.

\label{fig:images}



\section{Fast-Response Current Mode $^{6}$Li Glass Scintillator Detector}\label{s:detector}
 The current mode detector used in our apparatus was designed to be identical to one used by the TRIPLE Collaboration~\cite{Bowman90}. The reasons for this design choice were the similarities in neutron flux and the sensitivities achieved in the TRIPLE Collaboration's previous experiments~\cite{CDBowman1989, Yuan91}. The main goal of the design is to convert the neutron flux in a neutron-absorbing scintillator detector into an output current as a function of neutron time-of-flight. When the neutron rate is low, the detector can also resolve individual pulses in  pulse counting mode.

\subsection{Scintillator Characteristics}
The neutron detector shown in Fig. 1  contains a 13.3 cm diameter $\times$ 1 cm thick cylinder of Scintacor GS20 $^{6}$Li loaded glass optically coupled to a photomultiplier tube (PMT). The GS20 $^{6}$Li loaded glass has a density of $2.5$ g/cm$^{3}.$  The reaction n + $^{6}$Li $\rightarrow$ $^{7}$Li* $\rightarrow$ $^{4}$He + $^{3}$H + 4.8 MeV is used to detect neutrons. The GS20 glass is 6.6\% lithium by weight, enriched to 95\% $^{6}$Li. The 1 cm thickness of lithium glass gives a neutron absorption efficiency of 90\% for 1 eV neutrons. Due to the 1/$v_n$ dependence of the n + $^{6}$Li $\rightarrow$ $^{4}$He + $^{3}$H reaction cross section, the efficiency is lower at higher energies. The cross section of this reaction at 1 eV neutron energy is approximately 147 b~\cite{Harvey1979}. The scintillation light from neutron capture in GS20 glass has a fast 18 ns component, a slower 57 ns component, and a 98 ns rise time for the signal to rise from 10\% to 90\% of its full value. The scintillation light produced by the energy deposited by the  $^{4}$He and $^{3}$H ions is detected in the PMT.

An otherwise identical detector/PMT combination was constructed which uses glass depleted in $^{6}$Li so that it is very insensitive to neutrons but has an almost identical response to gammas. The attenuation of gammas in the materials used in the scintillator/PMT is low enough that one can place the $^{7}$Li-rich detector directly behind the $^{6}$Li-rich detector to measure and subtract out the signal from the gammas in the beam if needed. Based on the high signal/background ratio of our signal at neutron time of flights corresponding to the 1 eV region, we decided not to install the $^{7}$Li-rich backing detector for the double lanthanum measurements.  

\begin{figure}[H]
\centering
\includegraphics[width=0.9\linewidth]{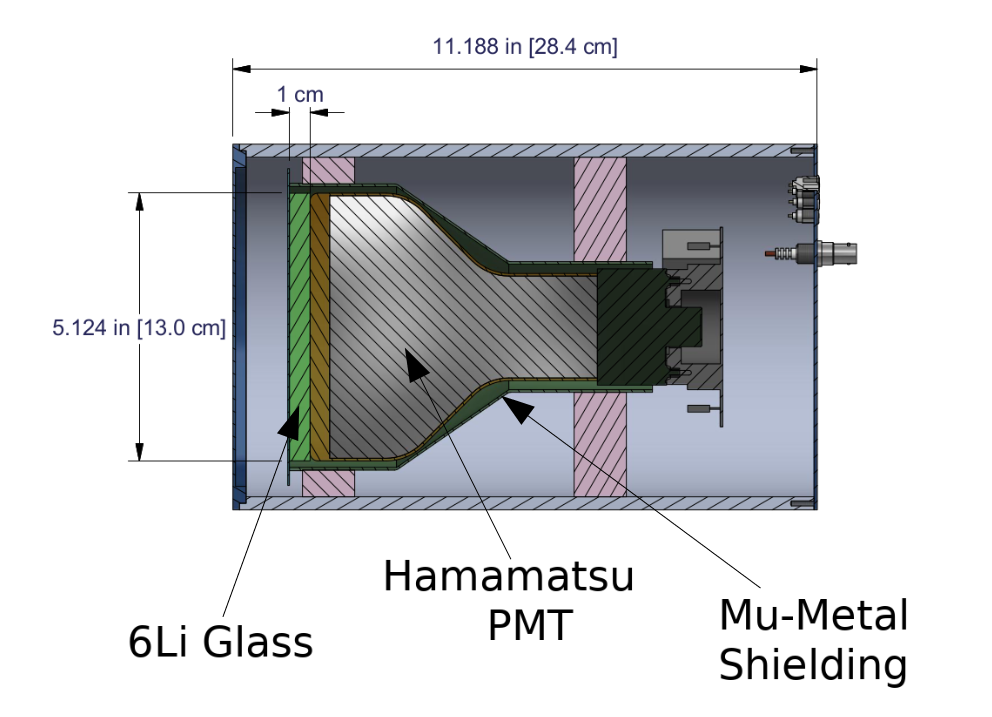}
\caption{\label{fig:Detector} Side view of the detector design. From left to right: an aluminum window 18 cm in diameter, the $^{6}$Li glass scintillator, the Hamamatsu R1513 PMT with its $\mu$-metal magnetic shield, the PCB, and the exit window with electrical feedthroughs. Not shown in this cutaway is the LED installed to produce pulsed light used for calibration purposes.}
\end{figure} 

\subsection{PMT and Analog Electronics}
\begin{figure}[H]
\centering
\includegraphics[width=1\linewidth]{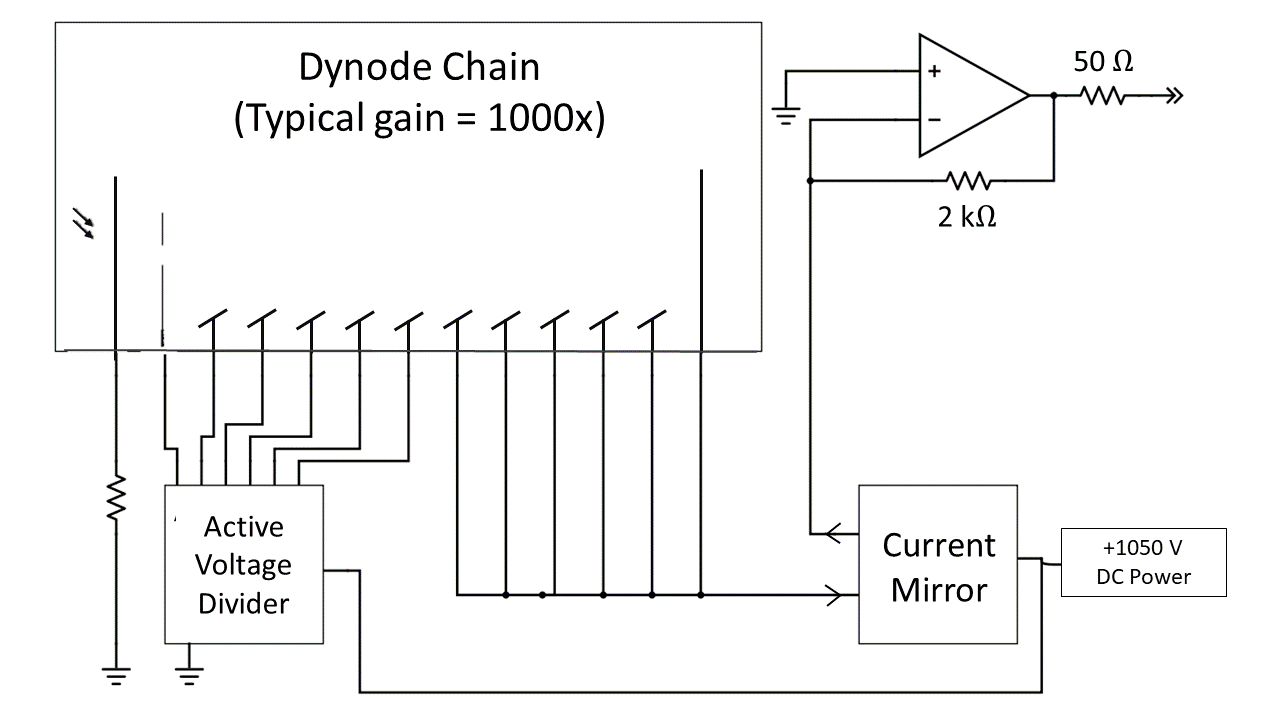}
\caption{\label{fig:PMT_Rev2} A simplified schematic of the PMT assembly.}
\label{fig:pmtcircuit}
\end{figure}

We chose a Hamamatsu R1513 PMT with a S-20 photocathode for its low photocathode resistivity and corresponding superior performance in current mode~\cite{Bowman90}. The maximum gain for this PMT is around 3.3 $\times 10^{5}$. A single photoelectron pulse at the anode of the PMT has a rise-time of around 7 ns. The PMT base circuitry is shown in Figure \ref{fig:pmtcircuit}. The photocathode current was specified to output 2 $\mu$A with a full scale of 2 mA. The full scale current requires a gain of 1000 within the dynode chain. This gain was achieved with a 1050 V input. The 2 $\mu$A current enables the detector to handle an instantaneous rate of 10$^{11}$ neutrons/s striking the detector given that there are typically 100 photoelectrons produced in the PMT when a single neutron is captured. The cathode within the PMT was grounded to ensure stable, low-noise operation of the PMT. The baseline level was designed to be 0.0 V which corresponded to 0 $\mu$A. A 2 $\mu$A current in the PMT produced an output voltage of -2.0 V. The PMT uses an active voltage divider so that the divider ratio is unaffected by load currents from the PMT. A light emitting diode was also included in the design of the PMT assembly so that testing could be performed with a internal LED. In addition to the high voltage supply needed to run the PMT, a +8V/-8V supply powers a buffering stage operational amplifier. All output signal cables are LEMO connections.

\section{Data Acquisition}\label{s:daq}
This setup was designed to measure the total neutron transmission through matched disks of \(^{139}\)La. Because of the high instantaneous flux rates needed for such a transmission measurement, it is not feasible to count individual neutron pulses; instead, the total integrated current output of the detector is measured. Our data acquisition system, like our detector, must be designed to handle the output signal in real time. 

We used a CAEN V1724 8-channel, 14-bit digitizer with a maximum sampling rate of 100 MS/s to record the signals from the \(^3\)He ion chamber, current-mode neutron detector, and 3 Faraday pickup loops (5 signals in total). The facility \(t_0\) signal from the 20~Hz Lujan Center accelerator proton current pickup coil was used to simultaneously trigger the spin-flipper controller and the data acquisition. Upon receiving a \(t_0\) trigger, 75000 2-byte samples were recorded for each of the 5 input voltage signals and stored in the onboard memory buffers where it was read out via an optical link to the data acquisition computer. In order to reduce the data file size, the respective waveforms were then decimated in software before being written to long-term storage. All further data manipulation and analysis was handled offline.

\subsection{Signal processing}
Higher-resolution spectroscopy of the 0.7~eV resonance was desired in order to increase the precision with which we could measure the Doppler-narrowed 0.7 eV resonance peak. To do so, each of the 5 signals were sampled every 10~ns by the V1724 digitizer and then decimated on-board by a factor of \(2^6=64\). After being read out by the CAEN board to the data acquisition desktop computer, the data underwent further decimation where it was structured into `windows,' i.e. different regions of the waveform were decimated by different factors in order to keep the data files at a reasonable size. The total data set for the Double Lanthanum experiment was $\sim4$ TB. If such a decimation scheme had not been implemented, the data set was expected to be well over 400 TB, which is unwieldy in terms of data storage and manipulation during the analysis process.

\subsection{Determination of Spin State}
\label{subsec:daq_faraday}
Extreme caution must be exercised when deciding how to record the spin state of the flipper in the datastream. If one chooses to measure a signal that correlates directly with the absolute spin state of the flipper (e.g. using a magnetic field probe to monitor the actual value of the \(\vec{B}\) field in real time), then it is possible that nonzero false asymmetries may creep into the data via insidious means such as electronic cross-talk between channels on the data acquisition modules. Although data acquisition modules are designed to minimize cross-talk, we did not want to risk such an occurrence in a high-statistics, precision experiment. Systematic ucertainties that effectively mimic a physics asymmetry completely compromise the validity of such an experiment.

To monitor the spin flipping process, each of the three coil sets (longitudinal, transverse, and shunt) had a respective lead wire threaded through its own pickup coil. Each pickup coil consisted of a toroidal solenoid wrapped around an iron core. The ferromagnetic core ensured that maximal magnetic flux was captured. Because the state of the flipper is determined by the direction of the current through the coils, any change in current will induce a voltage in the pickup coil. We measured these induced voltages to determine spin state transitions, from which the preceding and succeeding spin states could be determined. 

\begin{figure}[H]
\centering
\includegraphics[width=0.9\linewidth]{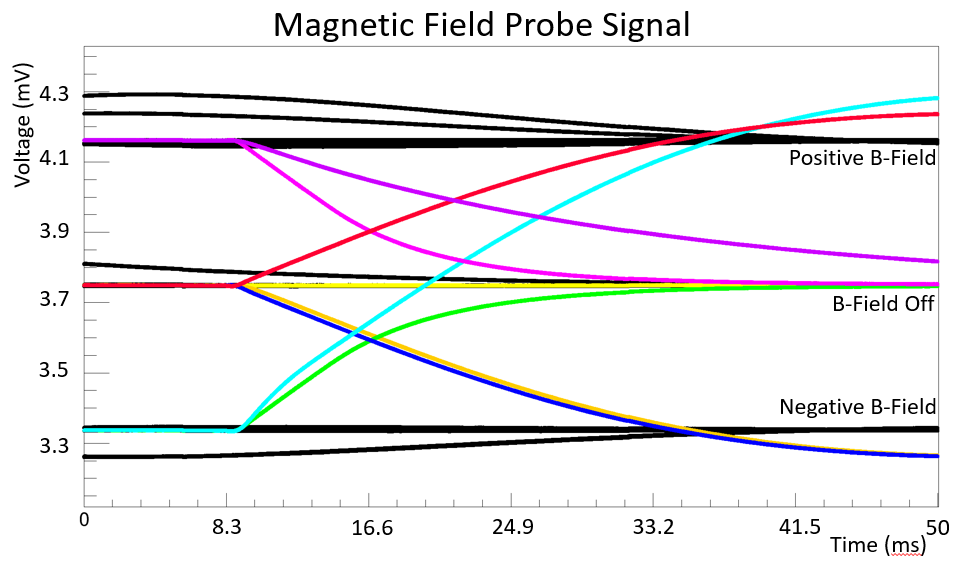}
\caption{\label{fig:transitions} This figure shows the pulses that were identified using the cuts made on the histograms of integrated voltage signals. Note all eight transitions are present, including the `no change in state' transition, shown in yellow.}
\end{figure}

Because the longitudinal spin transport field is always on, our spin flipper state can be defined by the state of the transverse field coil, of which there are three: off, no transverse field (0); on, positive transverse field (+); on, negative transverse field (-). For some early test runs, a Lakeshore 460 triple axis magnetic field probe was placed at the center of the spin flipper to record the absolute state of the transverse field to be used to test our sorting algorithm (this probe was later removed so that there was no chance of interference with our dataset). Figure \ref{fig:transitions} shows the overlaid field probe readings of a few thousand pulses. We can see that there are three stable magnetic field states, indicated by the three distinct levels in Figure \ref{fig:transitions}, as well as the transitions happening between them. We have developed a method to sort the spins by identifying the pickup coil voltage signature for each transition and used this information to tag each pulse as either \textit{flip} or \textit{no-flip}, described below.

\begin{figure}[H]
\centering
\includegraphics[width=\linewidth]{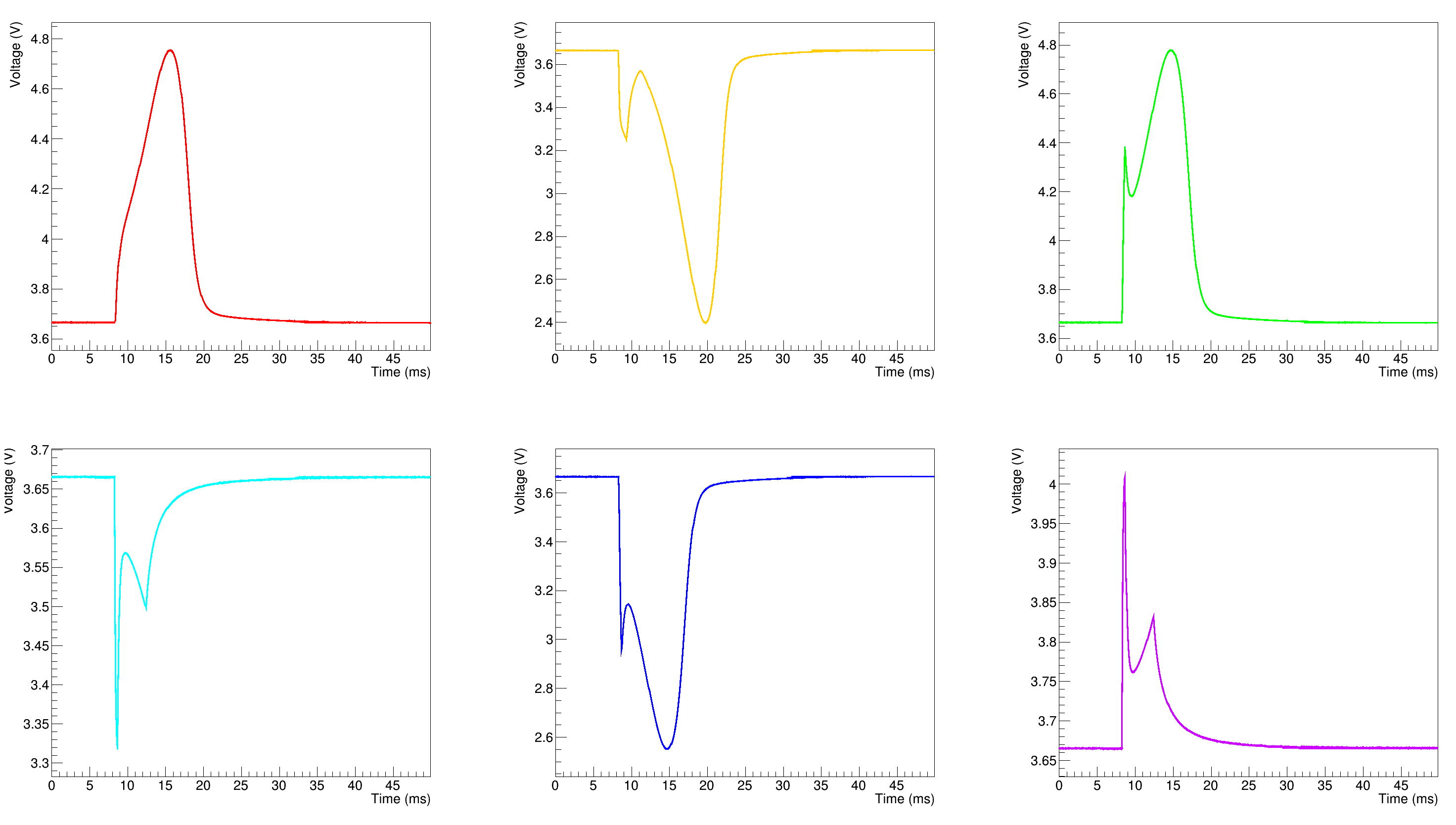}
\caption{Voltage signals were produced in the pickup coils by changing the currents flowing through the spin flipper coils. Here we see examples of 6 different spin state transitions.} 
\label{fig:voltages}
\end{figure}

\begin{figure}[H]
\centering
\includegraphics[width=0.9\linewidth]{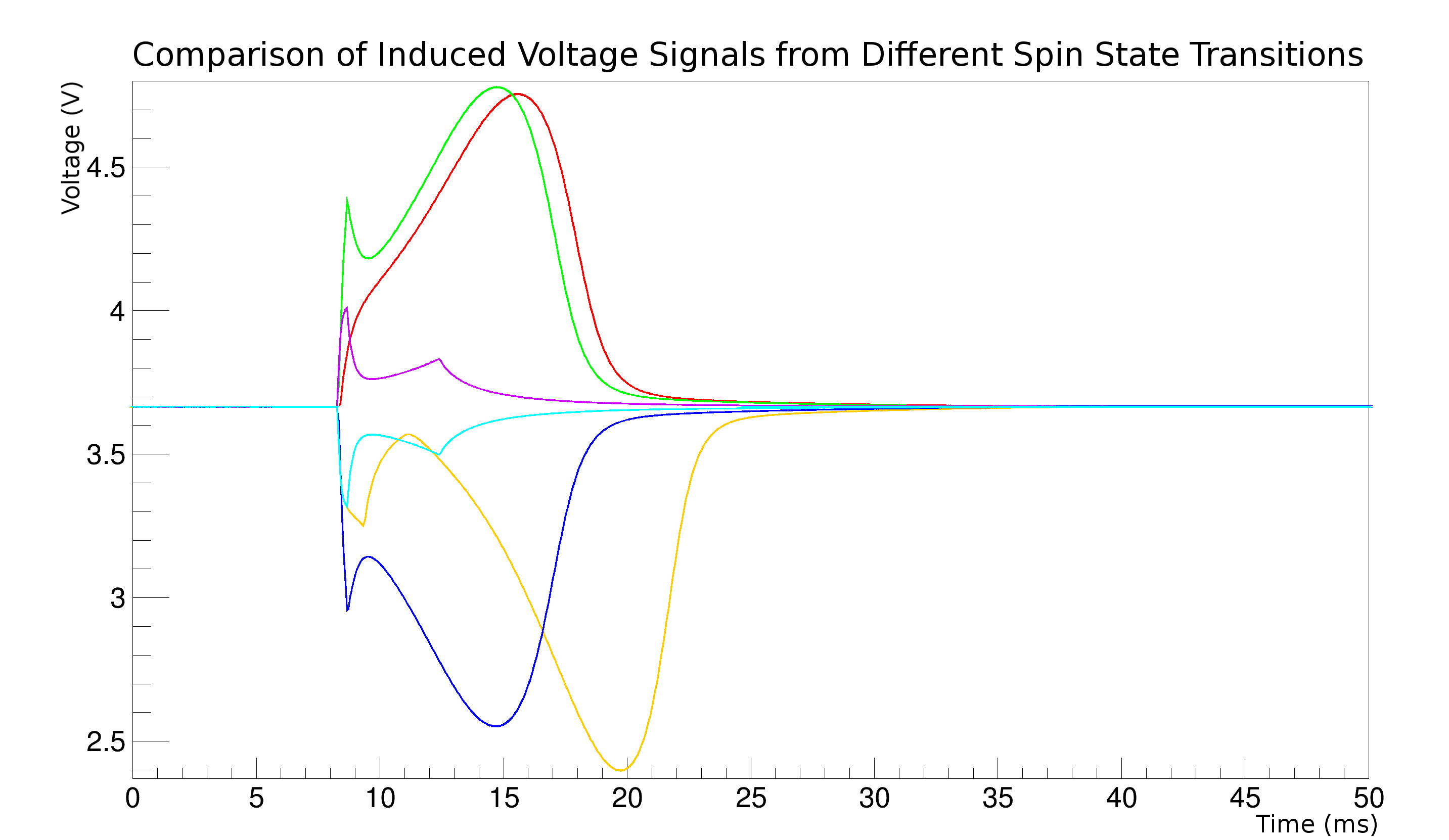}
\caption{Induced voltage signals for all 6 spin state transitions, superimposed to allow for comparison.}
\label{fig:voltage_all}
\end{figure}

Given our spin flip pattern of ($+00-0-+0$), we can see that we should have six unique transitions between states: \((+\rightarrow0), (0\rightarrow0), (0\rightarrow-), (-\rightarrow0)\), \((-\rightarrow+)\), and \((0\rightarrow+)\). Figure \ref{fig:voltages} shows a few examples of these voltage signatures that were produced when the spin flipper transitioned states, and Figure \ref{fig:voltage_all} shows these same traces superimposed on one another to show relative lineshapes and amplitudes. However, we have found that all eight transitions (the six unique transitions and the two degenerate \((0\rightarrow-)\) and \((+\rightarrow)\) transitions)

 between states can be uniquely identified by the pickup loop signature--this is due to the small differences in the placement of the pickup coils in relation to the spin flipper, allowing each to capture dissimilar enough magnetic flux changes to produce distinct voltage signatures.

\begin{figure}[H]
\centering
\includegraphics[width=0.9\linewidth]{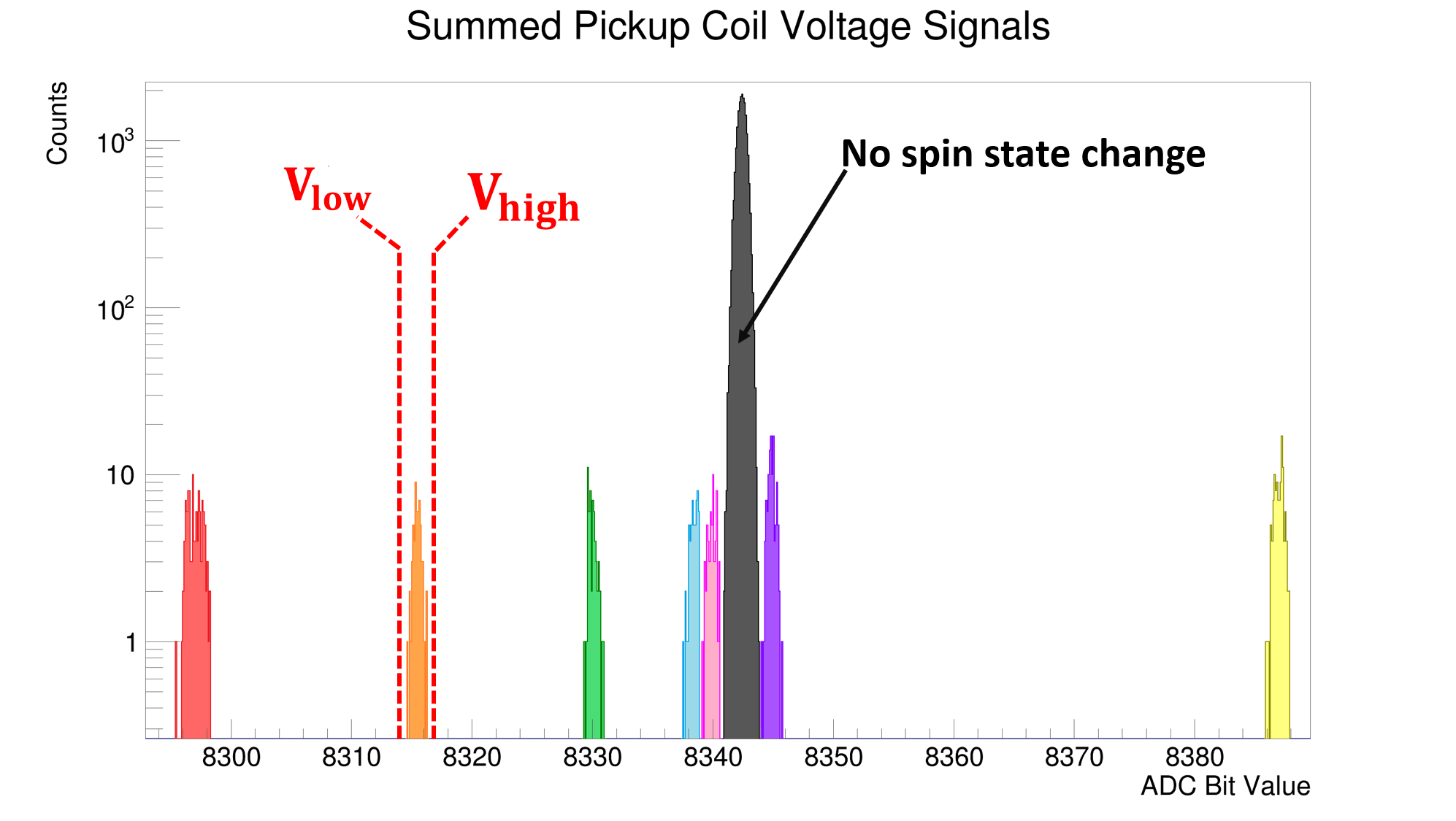}
\caption{\label{fig:sorting}A histogram showing the summation of each voltage signal produced by a spin state transition shows unique values for each of the states. These values were then used to sort pulses into their appropriate spin states.}
\end{figure}

To sort the data pulses by spin state, each voltage signal for a given pickup coil was integrated over the 50 ms pulse to return a single value for each pulse. These values were then histogrammed and clear peaks emerged for each spin state transition. By defining the upper and lower bounds for each peak on the histogram, we were able to determine and sort each spin state by understanding which transition was happening for each pulse and tagging the preceding spin states appropriately. Figure \ref{fig:sorting} shows a sample of some of the peaks, though it is worth noting that there were more than 8 peaks due to the summation of the pulses immediately preceding the initial transitional pulses causing small peaks very near the `no spin state change' peak.

\section{$^3$He Spin Filter Apparatus and Design}\label{s:3He}

A polarized $^3$He gas neutron spin-filter works by utilizing the spin dependent neutron absorption cross section of polarized $^3$He. Neutrons with their spins parallel to that of the $^3$He gas will be transmitted while neutrons with anti-parallel spins will be absorbed. A polarized $^{3}$He neutron spin filter of sufficient thickness and $^{3}$He polarization can polarize the neutron energies of interest between 0.1-10 eV with neutron polarization and transmission high enough to conduct several interesting experiments. The efficient removal of the antiparallel spins by absorption rather than scattering as in a neutron spin filter based on polarized proton scattering,  combined with the spatial uniformity of the $^{3}$He gas polarization in the cell,  makes the polarized neutron beam phase space highly uniform, This is an advantageous property for an eventual time reversal violation experiment as it suppresses possible sources of systematic error associated with neutron small angle scattering in the polarized target. The energy-dependent neutron transmission measurements possible at a pulsed spallation neutron source like LANSCE allow the neutron beam polarization produced by such a neutron spin filter to be determined to high accuracy by comparing the relative neutron transmission intensity for the $^{3}$He in the spin filter polarized versus unpolarized~\cite{Musgrave}. 
Polarized $^3$He gas is produced using spin-exchange optical pumping (SEOP)~\cite{Walker1997}. SEOP refers to the process of using optical pumping to polarize rubidium vapor and allowing the spin polarization of the Rb electrons to be transferred to $^{3}$He nuclei via spin exchange during gas-phase collisions. This spin-exchange is mediated by the hyperfine interaction between the Rb valence electron and the $^3$He nucleus. Below we include a brief description of the physics behind this process and our design of the system. Extensive references to previous work on polarized $^{3}$He neutron spin filters can be found in a recent review~\cite{Gentile2017}.  

\subsection{Spin Exchange Optical Pumping (SEOP) Theory}

\subsubsection{Rubidium Optical Pumping}
Optical pumping describes the process by which photons are used to redistribute the occupied states of some collection of atoms. Resonant absorption of light stimulates these states out of thermodynamic equilibrium toward a single desired state. Rubidium, as an alkali metal, is hydrogen-like because its outermost electron is shielded by complete inner shells. The Hamiltonian for the ground state of Rb in a static magnetic field $B_0 \hat{z}$ is:

\begin{equation}
\label{eq:larmor}
    {\cal H} = A_g\vec{I} \cdot \vec{S} + g_s \mu_B S_z B_0 \ - \frac{\mu_I}{I}I_z B_0
\end{equation}

where $\vec{I}$ is the Rb nuclear spin, $\vec{S}$ is the electron spin, $A_g$ is the isotropic magnetic dipole coupling coefficient, $g_s$ is the electron g-factor, $\mu_B$ is the Bohr magneton, and $\mu_I$ is the nuclear magnetic moment. The first term in the Hamiltonian is the hyperfine interaction between the nucleus and electron, while the second and third terms are the Zeeman interactions of the electron and nucleus with the external magnetic field. In the weak-field limit, the hyperfine interaction is larger than the Zeeman interaction, producing hyperfine splitting on the order of GHz and Zeeman splitting on the order of MHz. The total angular momentum of the atom is then $\vec{F} = \vec{I} + \vec{J}$ with its projection $m_F$ onto the the $z$-axis.

For rubidium, the ground state $^{2}S_{1/2}$ and excited state $^{2}P_{1/2}$ have an energy separation corresponding to a light wavelength of $\lambda = 794.8$ nm. These states are separated further into two $F$ states due to hyperfine splitting. Increasing the $B$-field further separates the $F$-states into $2F+1$ subdivisions. In the $^3$He gas cell there are two isotopes of Rubidium: $^{85}$Rb with nuclear spin $I=5/2$ and hyperfine levels $F=3,2$ and $^{87}$Rb with nuclear spin $I=3/2$ and hyperfine levels $F=2,1$. An unpolarized $794.8$ nm photon will cause a $^{2}S_{1/2} \rightarrow {}^{2}P_{1/2}$ transition according to the selection rules: $\Delta F = 0,\pm 1$ and $\Delta m_F = 0,\pm 1$. However, if said photon is circularly polarized with angular momentum pointing in the direction of the static magnetic field, this further restricts the allowed transition. The new selection rules are then: $\Delta F = 0,\pm 1$ and $\Delta m_F =\pm 1$. The excited state will then spontaneously decay back into the ground state by emission of a photon with arbitrary polarization and $\Delta m_F = 0,\pm 1$ selection rules. 
Given sufficient time, nearly all the Rb atoms will migrate to the highest $m_F$ value possible because there is no transition out of this state. This $m_F$ value corresponds to a situation where the electron and nuclear magnetic moments point in the same direction as the static magnetic field. The Rb atoms are therefore polarized and in the ground state. 


To improve the efficiency of this process a small amount of $\rm{N_2}$ gas is added to the cell. When an excited Rb atom spontaneously decays the emitted photon has a random polarization and can re-excite another Rb atom. Because the photon is not circularly polarized, this allows for $\Delta m_F = 0,\pm 1$ transitions. This in turn will slow down the polarization of the Rb atoms. The $N_2$ in the cell circumvents this bottleneck by allowing the Rb atoms to de-excite before they spontaneously decay. This relaxation is achieved by transferring the energy to the rotational and vibrational states of the $N_2$ molecules. This is made possible by the large quenching cross section of $5.8 \times 10^{-15}$ cm$^{-2}$ for $N_2$ gas~\cite{Hrycyshyn1970}. The relaxation takes place $\sim$10 times faster than the spontaneous decay.

\subsubsection{Spin-Exchange}
The next stage in polarization of $^3$He is the spin-exchange between the Rb valence electron and the $^3$He nucleus. From the perspective of the spins, the collision is dominated by two interactions:
\begin{equation} \label{eq:3He}
    V = \gamma \vec{N} \cdot \vec{S} + \alpha \vec{K} \cdot \vec{S}
\end{equation}
The first interaction is between the Rb valence electron spin $\vec{S}$ and the rotational angular momentum $\vec{N}$ of the Rb and $^3$He pair. The second interaction is the hyperfine coupling between the Rb valence electron spin $\vec{S}$ and the nuclear spin $\vec{K}$ of the $^3$He atom. The constants $\gamma$ and $\alpha$ are functions of the separation distance $R$ and determine the interaction strength. Spin relaxation of the Rb electron is from the spin-rotation interaction while the spin-exchange is caused by the hyperfine interaction. During a collision, the spin angular momentum of the Rb electron is transferred to the $^3$He nucleus. Because this spin exchange is a slow processes, acquiring maximally polarized $^3$He gas takes on the order of 10 hours. 

\subsubsection{Cell and Oven}
\begin{figure}[H]
\centering
\includegraphics[width=0.6\linewidth]{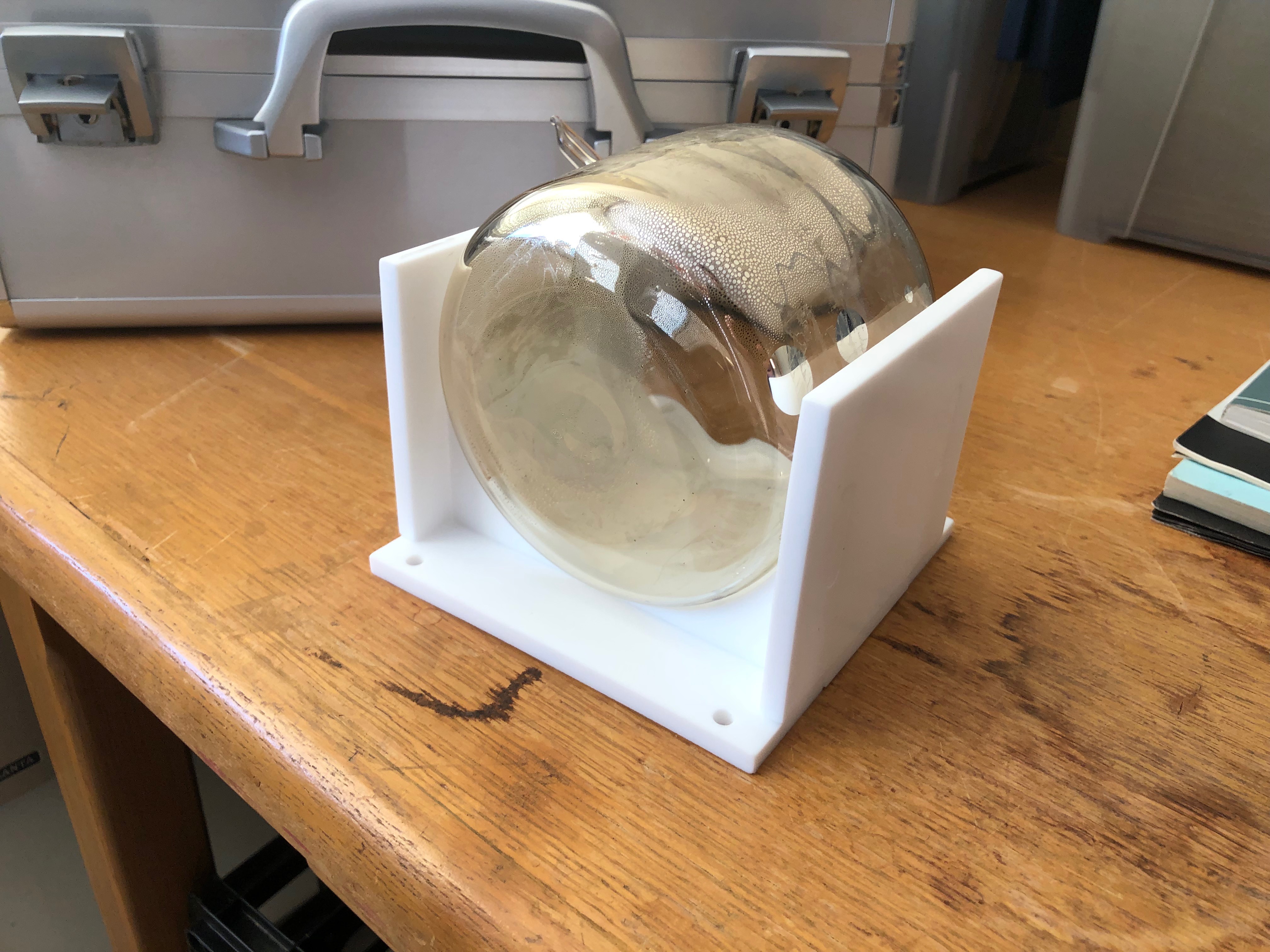}
\caption{\label{fig:3He_cell} $^3$He gas cell in its Teflon holder.}
\end{figure}

The $^3$He gas cell, shown in Figure \ref{fig:3He_cell}, is 12.1 cm in diameter and 10.8 cm thick in the longitudinal direction. The cell contains $^3$He gas at 2 bar pressure, 0.09 bar of N$_2$ gas, and solid Rb inside. To turn the solid Rb into a vapor, the cell needs to be heated to an ambient temperature of about 200 $^\circ$C. The cell is held in place with a Teflon holder and high temperature polyimide tape. Teflon is used because of its high heat-resistant properties and flexibility. Additionally, the Teflon holder supports two coils used for free induction decay (FID) NMR. The purpose of FID NMR is to measure the $^3$He polarization in a non-destructive way.

\begin{figure}[H]
\centering
\includegraphics[width=0.6\linewidth]{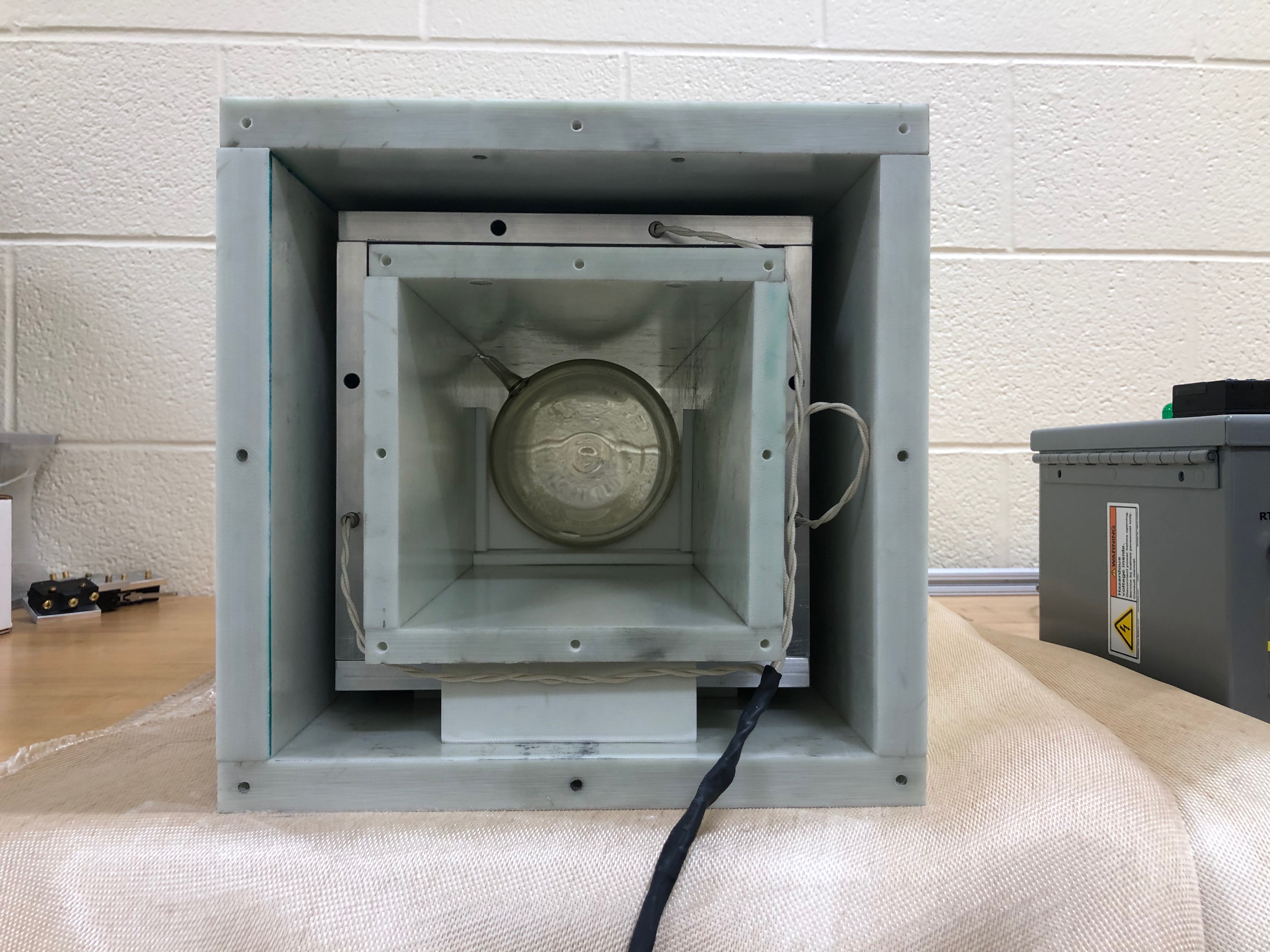}
\caption{\label{fig:oven_2} Picture of oven with $^3$He gas cell and cartridges heaters installed. }
\end{figure}

The oven used to heat the cell is a $26.7 \times 26.7 \times 63.5$ cm$^3$ rectangular box with 3 layers. The first layer is made of 1.27 cm Garolite ``G7" and supports the gas cell and adiabatic fast passage (AFP) NMR coil. The AFP NMR coil is used to flip the $^3$He polarization direction. The second layer is made of aluminum and holds six 7.6 cm electrical heating cartridges (Omega: CSH 103220). The third layer is also made of 1.9 cm G7, supports the first two layers and provides insulation. The end caps, also made of 1.27 cm G7, hold sapphire glass windows. G7 is a glass-silicon laminate, has a density of 4.57 g/cm$^3$, tensile strength of 20,000 PSI, and a maximum continuous operating temperature of 221$^\circ$C. These properties make it especially appropriate as a support and as an insulator for an oven with a target temperature of  200$^\circ$C. The purpose of the aluminum layer is to better distribute the heat created by the cartridge heaters. The outermost layer is supported by an adjustable support to better align the $^3$He gas cell to the neutron beam. Almost all oven components are made with non-magnetic material such as aluminum and brass. Figure \ref{fig:oven_2} shows the $^3$He cell situated inside of the oven.


\subsubsection{Laser Optics}
\begin{figure}[H]
\centering
\includegraphics[width=0.9\linewidth]{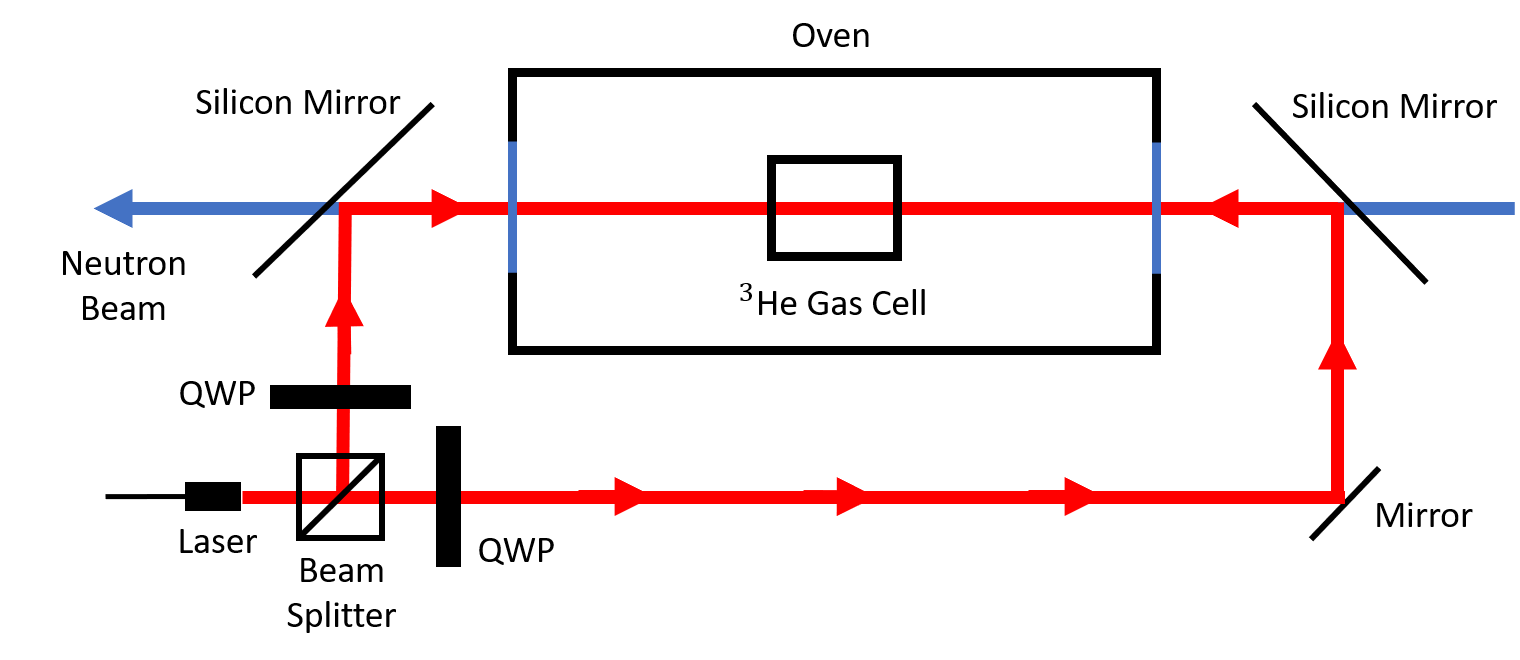}
\caption{\label{fig:optics_dia} Diagram showing the optics layout for creating circularly polarized light.}
\end{figure}

\begin{figure}[H]
\centering
\includegraphics[width=0.6\linewidth]{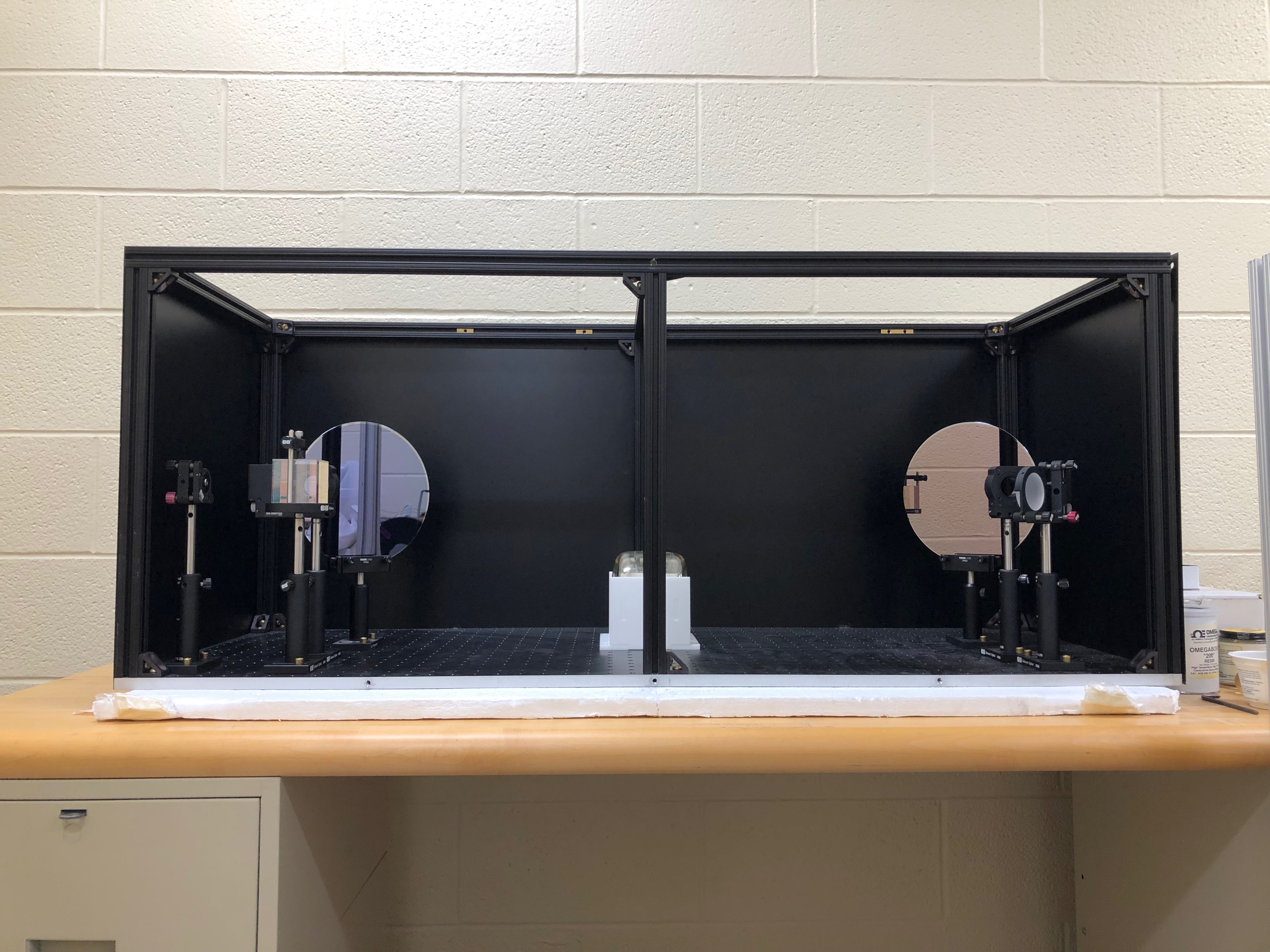}
\caption{\label{fig:optics} Picture of optics setup in black-anodized light tight box.}
\end{figure}

To excite the Rb atoms, a 50~W fiber-coupled 795 nm laser is used. Circularly polarized light is created by passing linearly polarized light through a quarter wave plate at a 45$^\circ$ angle. First, the laser fiber is coupled to a beam collimator (ThorLabs: F220SMA-780) that focuses and reduces the spread of the beam. The laser is then incident on a polarizing cube beam-splitter (Edmund Optics: 49-872) that linearly polarizes and splits the beam 90$^\circ$ into two paths. Both paths are then incident on a 2.54 cm diameter, 795 nm quarter wave plate (QWP). Each QWP is mounted on a precision motorized rotation device (ThorLabs: K10CR1). These devices allow for remote flipping of the direction of circular polarization by rotating the QWPs by 90$^\circ$. One path is then reflected by 90$^\circ$ using a 50 mm diameter dielectric mirror (Edmund Optics: 33-189). Finally, both paths are reflected in the oven with large 8 inch diameter dielectric coated silicon mirrors. The mirrors reflect the laser light, but are nearly transparent to neutrons. The laser light then passes through the sapphire glass windows and into the the $^3$He gas cell. Figure \ref{fig:optics_dia} shows a schematic of the optical layout. All optical components and oven are mounted on to a $48.2 \times 119.3 \times 1.27$ cm ($19 \times 47 \times 0.5$ inch) optical breadboard. Additionally, the breadboard and optics are enclosed in a black-anodized light tight box, shown in Figure \ref{fig:optics}. All optical components are made as non-magnetic as possible by using brass and aluminum parts.

\subsubsection{$\mu$-Metal Solenoid and Support}
\begin{figure}[H]
\centering
\includegraphics[width=0.6\linewidth]{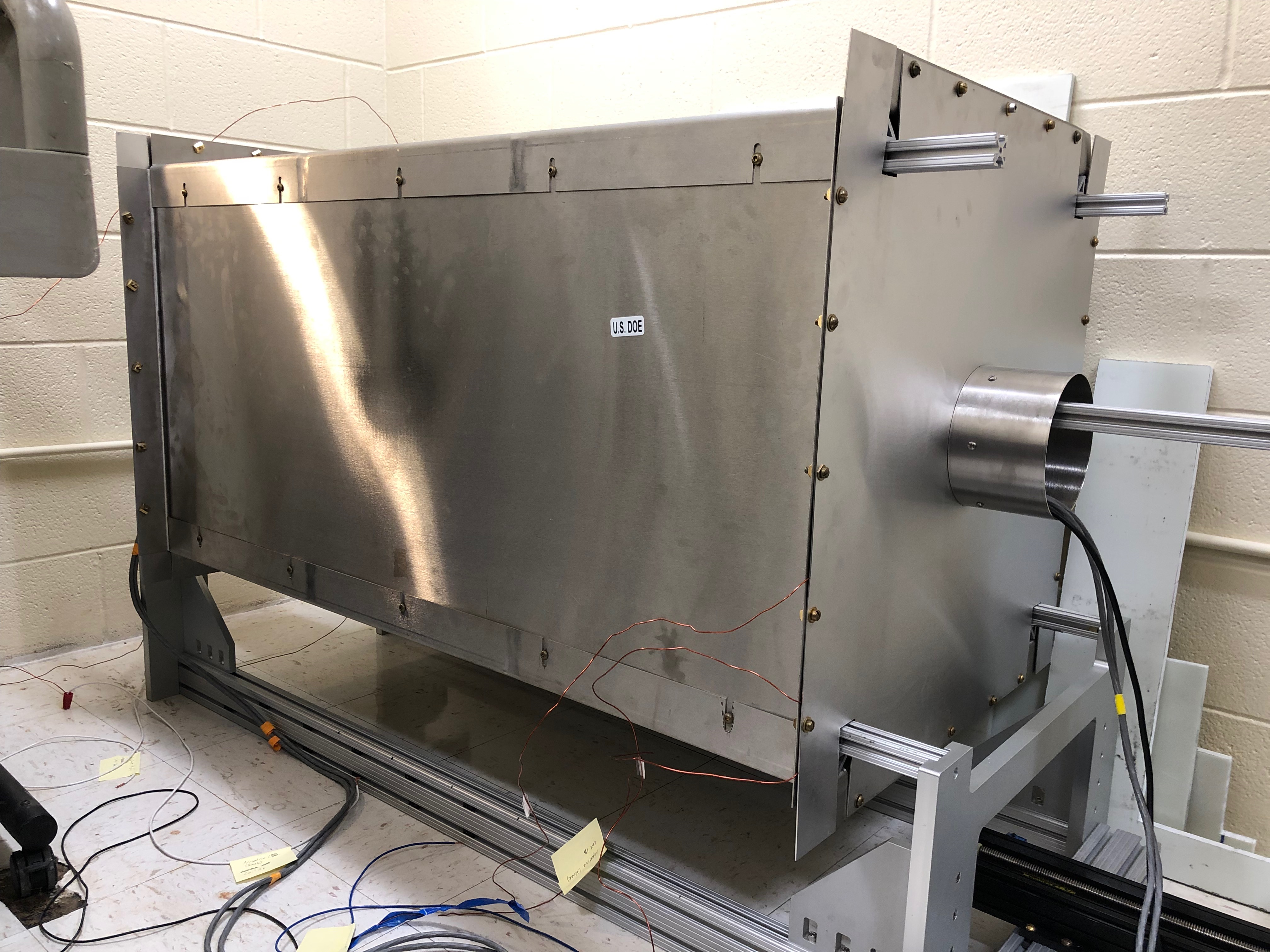}
\caption{\label{fig:mu_shield} Picture of $\mu$-metal shielded solenoid.}
\end{figure}

A $\mu$-metal shielded solenoid provides the uniform magnetic field required for the SEOP process. The $\mu$-metal shielded solenoid is approximately 61~cm $\times$ 61~cm $\times$ 122~cm  in size. The internal structure of of the $\mu$-metal shield is made of an 8020 frame of 2.5~cm $\times$2.5~cm profiles. The light tight box (with the optical components and oven mounted to the optical bread board) slides into the $\mu$-metal shield on a Teflon sheet. Figure \ref{fig:mu_shield} shows the shieldhouse that surrounds the solenoid, light tight box, and oven assembly. The $\mu$-metal shield is critical due to the $^3$He spin-filter's close proximity to the adiabatic spin-flipper. Because a very small magnetic field gradient is needed over the length of the $^3$He gas cell, it is crucial to shield it from external magnetic fields; $\mu$-metal works well as a magnetic shield due to its high magnetic permeability. Stray, slowly-varying magnetic fields are distributed around the shielding and away from the center where the $^3$He cell is located. The solenoid enclosed in the $\mu$-metal shield produces a magnetic field along the neutron beam direction that holds the polarization of $^3$He atoms. In the central 7 cm $\times$ 16 cm region where the $^3$He cell is located, the magnetic field has an average gradient $\langle\frac{\nabla B_x}{B_i}\rangle = 5.55\times10^{-4}\pm2.66\times10^{-4}$ cm$^{-1}$ along the longitudinal polarization direction, and an average gradient $\langle\frac{\nabla B_y}{B_i}\rangle = 6.03\times10^{-4}\pm1.68\times10^{-4}$ cm$^{-1}$ in the transverse direction. Based on these measurements, the expected relaxation time of the $^3$He is $~100-200$ hours. The solenoid has a resistance of $R=25.8$ $\Omega$, and is driven at 1.00 A with a current stability of 0.1$\%$. At this current, the solenoid produces a holding field of 9.26 gauss at its center which, along with an 30 kHz NMR pickup coil, is tuned to the 3.24 kHz/gauss gyromagnetic ratio of $^3$ He\cite{Flowers1993}. The NMR signal is then used to determine the polarization of the $^3$He gas.

Due to the modular nature of the apparatus, this polarized $^{3}$He neutron spin filter can be installed at the location of the first cryostat (upon removal of the cryostat from the apparatus). It can then be operated continuously on-line during measurements, as was successfully done several years ago using a similar apparatus~\cite{Chupp2007} for the first phase of the NPDGamma neutron-proton weak interaction experiment~\cite{Gericke2011}.

\section{Summary and Outlook}\label{s:summary}

We have described a flexible, rotatable cryogenic apparatus for the measurement of parity violation in the transmission of longitudinally polarized neutrons in neutron-nucleus resonances in the 0.1-10 eV energy range at a short-pulsed spallation neutron source. The apparatus employs beam monitors and transmission detectors combining both (n, $\gamma$) discrimination and current mode operation, a broadband neutron spin flipper with a flipping mode and sequence designed to suppress systematic errors, and a neutron polarizer based on spin-exchange optical pumping of $^{3}$He. The apparatus design employs several features used in previous neutron parity violation experiments at LANSCE, especially from work performed by the TRIPLE and NPDGamma collaborations and referred to in the text.    

The apparatus that we have described also possesses sufficient flexibility to accommodate several modifications that can greatly extend the range of possible experiments beyond the Double Lanthanum experiment that we concentrated on as an example in this paper. The first major addition to the apparatus we envision for the measurement of P-odd transmission asymmetries in other nuclei beyond $^{139}$La is the addition of the polarized $^{3}$He neutron spin filter based on spin-exchange optical pumping (described in Section 8) and which was first demonstrated at LANSCE~\cite{Coulter}.  The experimental hutch on the FP12 beamline possesses an access path through the ceiling near the present location of the downstream refrigerator that we plan to use for the operation of cryostats, which require safe venting of cryogenic liquids into a storage volume on top of the hutch, as done in the past for liquid parahydrogen targets operated on FP12~\cite{Santra2010, Palos2011}. In particular, this implementation would allow us to conduct a new measurement of the parity-odd asymmetry in the 3.2 eV p-wave resonance in $^{131}$Xe, which is one of the leading candidates for a sensitive time reversal test, using a solid xenon target. Liquid helium supply and venting for the dilution refrigerator required for the polarized nuclear targets to be used in the time reversal test can also be provided through this path. 

Much care was taken in designing this apparatus to minimize any systematic effects that would affect the uncertainty associated with the P-violation measurements. Because this apparatus was designed to be modified to accommodate several nuclear species, each measurement will have its own associated systematic effects. The most dominant systematic effect associated with the design of the apparatus itself lies in how well one can determine the neutron spin flip efficiency. In this paper, we present a method to calculate this efficiency using flux monitor measurements of the beam intensity convoluted with the image plate beam profiles and measurements of the $B$-field combined with a Monte Carlo calculation of the neutron intensity-weighted spin flip efficiency integrated over the beam cross sectional area. In this case, the dominant uncertainty will lie in how well one can reconstruct the $B$-field produced by the spin flipper coils, i.e. the spatial resolution of the $B$-field maps, which is on the order of a few parts in 10$^{3}$. If this accuracy does not suffice one can install a $^3$He polarizer/ analyzer  pair and measure the neutron spin flip efficiency directly using the method described by Musgrave et al. in \cite{Musgrave}.

The beamline components can be translated upstream far enough to accept gamma detector arrays downstream of the target to conduct parity violation measurements using (n, $\gamma$) reactions~\cite{Seestrom1999} as well as in neutron transmission measurements. The addition of a gamma detector array to this apparatus would enable one to efficiently search for large P-odd asymmetries between 0.1-10 eV in the several isotopes of heavy nuclei which have never been measured. The TRIPLE collaboration concentrated its choice of nuclei for measurement near the peaks of the 3p and 4p p-wave strength functions in an attempt to maximize the number of p-wave resonance per unit energy interval in the isolated resonance eV-keV regime for their statistical analysis of parity violation in heavy nuclei. Since that work was conducted, improved data on the p-wave strength function~\cite{Mughabghab2018a} has resolved an additional small peak near A=160, which is split off from the 4p giant resonance due to nuclear deformation effects as predicted long ago~\cite{Buck1962}, whereas the spherical optical model by contrast predicts a minimum at A=160. It therefore may be worthwhile to search in this A range to see if one can discover more p-wave resonances that amplify parity violation. One could also search for large parity-odd effects in the (n, $\gamma$) channel for nuclei that undergo neutron-induced fission and try to confirm and extend the many large P-odd asymmetries already measured in the fission channel~\cite{Petukhov1979, Borovikova1979, Borovikova1980, Petukhov1980, Danilian1980, Sushkov1982, Bunakov1983, Alexandrovich1984a, Alexandrovich1984b, Alexandrovich1987, Petrov1989, Alexandrovich1994}. In particular one could measure the P-odd asymmetries in the (n, $\gamma$) channel in very low energy resonances in $^{235}$U~\cite{Alfimenkov1999, Alfimenkov2000}, which is a nucleus that can be safely handled and supplied at the LANSCE facility. There is an already-proven method for nuclear polarization of $^{235}$U using the very large hyperfine fields produced at the nucleus in the low-temperature ferromagnetic phase of uranium sulfide. This phenomenon was exploited in the past to determine the spin assignments in $^{235}$U using spin-dependent polarized neutron transmission through polarized $^{235}$U ~\cite{Moore1978}. The nonmagnetic composition of the apparatus also lends itself to the use of polarized xenon targets produced by a spin-exchange optical pumping process very similar to that described above for $^{3}$He.  The 10 cm x 10 cm cross section, m=3 supermirror neutron guide presently installed upstream of the LANSCE FP12 hutch~\cite{Seo2004} could be replaced in the future with other types of neutron guides that can increase the fraction of neutrons transmitted to the apparatus near 1 eV. The dimensions of the FP12 hutch transverse to the neutron beam direction are wide enough to allow for the complete rotation of the relevant components of the apparatus by 180 degrees as envisioned in one of the modes of realization for a precision test of time reversal invariance~\cite{Bowman2014}. According to the results of MCNP simulations of the new configuration of the LANSCE target/moderator/reflector system scheduled to be installed in early 2020~\cite{Zavorka2018} we do not expect the projected changes, including the removal of the liquid hydrogen slow neutron moderator, to significantly influence the beam characteristics in our energy range of interest between 1-10 eV and therefore the performance of the components of the apparatus presented above should remain essentially unchanged after this upgrade.  

\section{Acknowledgements}\label{s:acknowledgements}

This material is based upon work supported by the U.S. Department of Energy, Office of Science, Office of Nuclear Physics, under Award Number DE-SC-0014622. D. Schaper would also like to acknowledge support by the NSF Graduate Research Fellowship Program under Grant Number 1247392. The work of C. Auton, J. Curole, J. Doskow, W. Fox, H. Lu, W. M. Snow. K. Steffen, J Vanderwerp, B. Short, and G. Visser was supported by NSF PHY-1614545, NSF PHY-1913789, and the Indiana University Center for Spacetime Symmetries. J. Curole also acknowledges support from the GAANN fellowship program of the US Department of Education and the Department of Energy SCGSR program. L. Cole and D. Olivera would like to acknowledge funding from the KY-NSF EPSCoR Research Scholar Program. B.M. Goodson acknowledges funding from NSF (CHE-1905341), DoD (W81XWH-15-1-0272), and a Cottrell SEED Award from the Research Corporation for Science Advancement. We  gratefully  acknowledge  the  local  support  of the  LANSCE  neutron  facility  at  Los  Alamos  National Lab where this measurement was performed. Test measurements for the current mode detector electronics used for this work were conducted in part on the NOBORU instrument at the JPARC MLF through proposal 2016B0021, Characterization of High counting rate Epithermal Neutron Detectors



\end{document}